\documentclass[twocolumn]{aastex62}
\submitjournal{ApJ}

\shorttitle{whistler waves at 1AU}
\shortauthors{Tong et al.}
\begin{document}

\title{Statistical Study of Whistler Waves in the Solar Wind at 1 AU}

\correspondingauthor{Yuguang Tong}
\email{ygtong@berkeley.edu}

\author{Yuguang Tong}
\affil{Space Sciences Laboratory, University of California, Berkeley, CA 94720}
\affil{Physics Department, University of California, Berkeley, CA 94720}

\author{Ivan Y. Vasko}
\affil{Space Sciences Laboratory, University of California, Berkeley, CA 94720}
\affil{Space Research Institute of Russian Academy of Sciences, Moscow, Russia}

\author{Anton V. Artemyev}
\affil{Institute of Geophysics and Planetary Sciences, University of California, Los Angeles, USA}2
\affil{Space Research Institute of Russian Academy of Sciences, Moscow, Russia}

\author{Stuart D. Bale}
\affil{Space Sciences Laboratory, University of California, Berkeley, CA 94720}
\affil{Physics Department, University of California, Berkeley, CA 94720}

\author{Forrest S. Mozer}
\affil{Space Sciences Laboratory, University of California, Berkeley, CA 94720}




\begin{abstract}
Whistler waves are intermittently present in the solar wind, while their origin and effects are not entirely understood. We present a statistical analysis of magnetic field fluctuations in the whistler frequency range (above 16 Hz) based on about 801,500 magnetic field spectra measured over three years aboard ARTEMIS spacecraft in the pristine solar wind. About 13,700 spectra (30 hours in total) with intense magnetic field fluctuations satisfy the interpretation in terms of quasi-parallel whistler waves. We provide estimates of the whistler wave occurrence probability, amplitudes, frequencies and bandwidths. The occurrence probability of whistler waves is shown to strongly depend on the electron temperature anisotropy. The whistler waves amplitudes are in the range from about 0.01 to 0.1 nT and typically below 0.02 of the background magnetic field. The frequencies of the whistler waves are shown to be below an upper bound that is dependent on $\beta_{e}$. The correlations established between the whistler wave properties and local macroscopic plasma parameters suggest that the observed whistler waves can be generated in local plasmas by the whistler heat flux instability. The whistler wave amplitudes are typically small, which questions the hypothesis that quasi-parallel whistler waves are capable to regulate the electron heat flux in the solar wind. We show that the observed whistler waves have sufficiently wide bandwidths and small amplitudes, so that effects of the whistler waves on electrons can be addressed in the frame of the quasi-linear theory.
\end{abstract}



\section{Introduction \label{sec1}}
 
Whistler waves, electromagnetic emissions between ion and electron cyclotron frequencies, are potentially regulating several fundamental processes in the collisionless or weakly-collisional solar wind. In particular, spacecraft observations of the electron heat flux values below a threshold dependent on $\beta_{e}$ were interpreted in terms of the heat flux regulation by the whistler heat flux instability \citep{Feldman75,Feldman76,Scime94,Gary1999b,Tong18} and whistler fan instability \citep{Vasko2019a}. The observed radial evolution of the  angular width of suprathermal field-aligned electron population (strahl electrons) in the solar wind \citep[e.g.,][]{Hammond96,Graham17} requires pitch-angle scattering that can be potentially provided by whistler waves \citep[][]{Vocks05,Shevchenko10,Vocks12,Kajdic:2016a,Vasko2019a}. Whistler waves may also suppress the electron heat flux in collisionless or weakly-collisional astrophysical plasma \citep{Pistinner1998a,Gary2000a,Roberg-Clark:2016,Roberg-Clark:2018b,komarov_2018}\replaced{ as required, e.g., by}{. The necessity  of a heat flux suppression mechanism is suggested by } observations of the temperature \replaced{distribution}{profile} of hot gases in galaxy clusters \citep[e.g.,][]{Cowie77,Meiksin86,Zakamska&Narayan03,Wagh14,Fang18}. The understanding of whistler wave origins and effects requires statistical analysis of whistler wave occurrence and properties in the solar wind.

The magnetic field fluctuations with power-law spectra in various frequency ranges are persistently observed in the solar wind and referred to as turbulence  \citep[see, e.g.,][for review]{Bruno13}. Early studies associate the magnetic field turbulence in the whistler frequency range with whistler waves, their power was shown to decrease with increasing radial distance from the Sun and enhance around interplanetary shocks and high-speed stream interfaces \citep[e.g.,][]{Beinroth81,Coroniti82,LengyelFrey96,Lin1998a}. However, later studies show that the whistler frequency range of the magnetic field turbulence is dominated by kinetic-Alfv\'en and slow ion-acoustic waves Doppler-shifted into the whistler frequency range \citep[e.g.,][]{Bale05,Salem12,Chen13,Lacombe17}. The whistler wave contribution to the magnetic field turbulence spectrum is still under debate \citep[e.g.,][]{Gary15,Narita16,Kellogg18}.

The modern spacecraft measurements have recently shown that whistler waves are intermittently present in the pristine (not disturbed by shocks or the Earth's foreshock) solar wind \citep{Lacombe14,Stansby16,Tong2019a}. Whistler waves have been identified by a local peak superimposed on a power-law spectrum of the magnetic field turbulence background. Therefore, these whistler waves should be produced by kinetic instabilities (free energy in the plasma), rather than by the turbulence cascade \citep[see][for discussion]{Gary15}. In addition to the pristine solar wind, whistler waves have been reported around interplanetary shock waves \citep[e.g.,][]{Breneman10,Wilson13} and in the Earth's foreshock \citep[e.g.,][]{Hoppe80,Zhang:1998}.

The focus of this paper is the statistical analysis of whistler waves produced by kinetic instabilities in the pristine solar wind. The detailed analysis of whistler waves in the pristine solar wind has become possible only recently due to simultaneous wave and particle measurements aboard Cluster and ARTEMIS spacecraft \citep{Lacombe14,Stansby16,Tong2019a}. In contrast to WIND and Stereo spacecraft, wave measurements aboard Cluster and ARTEMIS are available almost continuously, rather than triggered by high-amplitude events, which typically occur around interplanetary shocks \citep[e.g.,][]{Breneman10,Wilson13}. \cite{Lacombe14} have selected about twenty 10-minute intervals with whistler wave activity observed aboard Cluster in the pristine solar wind. The analysis of the magnetic field cross-spectra has shown that whistler waves propagate quasi-parallel to the background magnetic field. The simultaneous measurements of the electron heat flux have been presented to argue that the whistler waves are produced by the whistler heat flux instability (WHFI) \citep[see, e.g.,][for the WHFI theory]{Gary1994a}. \cite{Stansby16} have selected several 10-minute intervals of ARTEMIS measurements to test the whistler wave dispersion relation in dependence on $\beta_{e}$. \cite{Tong2019a} have carried out a detailed analysis of wave and particle measurements for \cite{Stansby16} events and demonstrated that the whistler waves were produced locally on a time scales of seconds and indeed by the WHFI. The analysis by \cite{Tong2019a} has proved that the WHFI may indeed operate in the solar wind and clearly demonstrated the critical role of the electron temperature anisotropy: the parallel temperature anisotropy may quench the WHFI instability, while the perpendicular temperature anisotropy favors the instability onset.  

In spite of some recent progress, the parameters controlling the occurrence and properties of whistler waves in the solar wind have not been considered on a statistical basis. In this paper we present analysis of several hundred days of ARTEMIS observations in the solar wind \citep[two spacecraft orbiting the Moon, see][for details]{Angelopoulos2011a}. The whistler wave selection produced a dataset of about 13,700 whistler wave spectra ($>30$ hours \added{in total}) in the pristine solar wind that is the most representative dataset up to date. The paper is organized as follows. We describe instrument characteristics, methodology and data selection criteria in Section \ref{sec2}. The results of the statistical study are presented in Sections \ref{sec3}, \ref{sec4} and \ref{sec5}. We discuss the statistical results in light of whistler wave generation mechanism, electron heat flux regulation and recent particle-in-cell simulations in Section \ref{sec6}. The conclusions are summarized in Section \ref{sec7}.

\section{Data and Methodology \label{sec2}}

We use ARTEMIS spacecraft measurements from 2011 to 2013 and select observations in the pristine solar wind, that is excluding the Earth's foreshock and the lunar wake. The Search Coil Magnetometer instrument provides Fast Fourier Transform (FFT) magnetic field spectra with 8s cadence and covers 64 piecewise linearly-spaced frequency channels between 8 to 4096 Hz \citep{Roux08}. We use the spectral power density SPD$_{\perp}$ of the magnetic field in the spacecraft spin plane (almost ecliptic plane), the spectral power density SPD$_{||}$ of the magnetic field component along the spin axis (almost perpendicular to the ecliptic plane) and, the total spectral power density $\mathrm{SPD}=\mathrm {SPD}_{||}+2\;\mathrm{SPD}_{\perp}$. The Flux Gate Magnetometer (FGM) provides the quasi-static magnetic field measurements at 4 vectors per second \citep{Auster08}, which we downsample by averaging to 8s cadence of the magnetic field spectra. The electron velocity distribution function (VDF) is measured every 3s by the Electrostatic Analyzer \citep{McFadden08}, and transmitted to the ground every 3 or 96s depending on the telemetry mode. We use the ground calibrated particle moments (density, bulk velocity and temperatures)\added{\footnote{Ground-calibrated particle moments are accessed via two data products,  \texttt{THB\_L2\_ESA} and \texttt{THC\_L2\_ESA} which can be found in \texttt{https://cdaweb.gsfc.nasa.gov/}.}} and the electron heat flux parallel to the magnetic field computed by integrating the electron VDF\added{\footnote{The electron VDF is accessed from \texttt{http://themis.ssl.berkeley.edu/data/themis/} and then processed by the open-source SPEDAS software \citep{Angelopoulos19}.}}
\begin{equation}
    q_e = \frac{1}{2} m_{e}\int (v_{||}-\left<v_{||}\right>)\;({\mathbf v}-\left<{\mathbf v}\right>)^2 \;{\mathrm{VDF}}({\mathbf v})\;d{\mathbf v}
    \label{eq:eq1}
\end{equation}
where $m_{e}$ is the electron mass, $v_{||}$ is the electron velocity parallel to the magnetic field and $\left<{\mathbf v}\right>$ is the electron bulk velocity. The particle moments available at 96s are upsampled to 8s cadence of the magnetic field spectra via the linear interpolation. In total we have analyzed 801,527 magnetic field spectra, spanning 1,803 hours and 359 days in 2011-2013 \replaced{(listed in the supplementary material)}{\footnote{The data intervals are provided in \texttt{https://doi.org/10.5281/zenodo.2652949}}}. In the rest of this paper, we will refer to each magnetic field spectrum as an independent event. Note that we did not filter out interplanetary shocks, but looking through the list of interplanetary shocks observed on Wind\replaced{(www.cfa.harvard.edu/shocks/wi\_data)}{\footnote{\texttt{www.cfa.harvard.edu/shocks/wi\_data}}}, we found only \replaced{several common days}{several days in our dataset with listed shocks}. Therefore, our dataset is dominated by observations in the pristine solar wind. In what follows, we clarify criteria for whistler wave selection and demonstrate the data analysis techniques.

\begin{figure*}
    \centering
    \includegraphics[width=\linewidth]{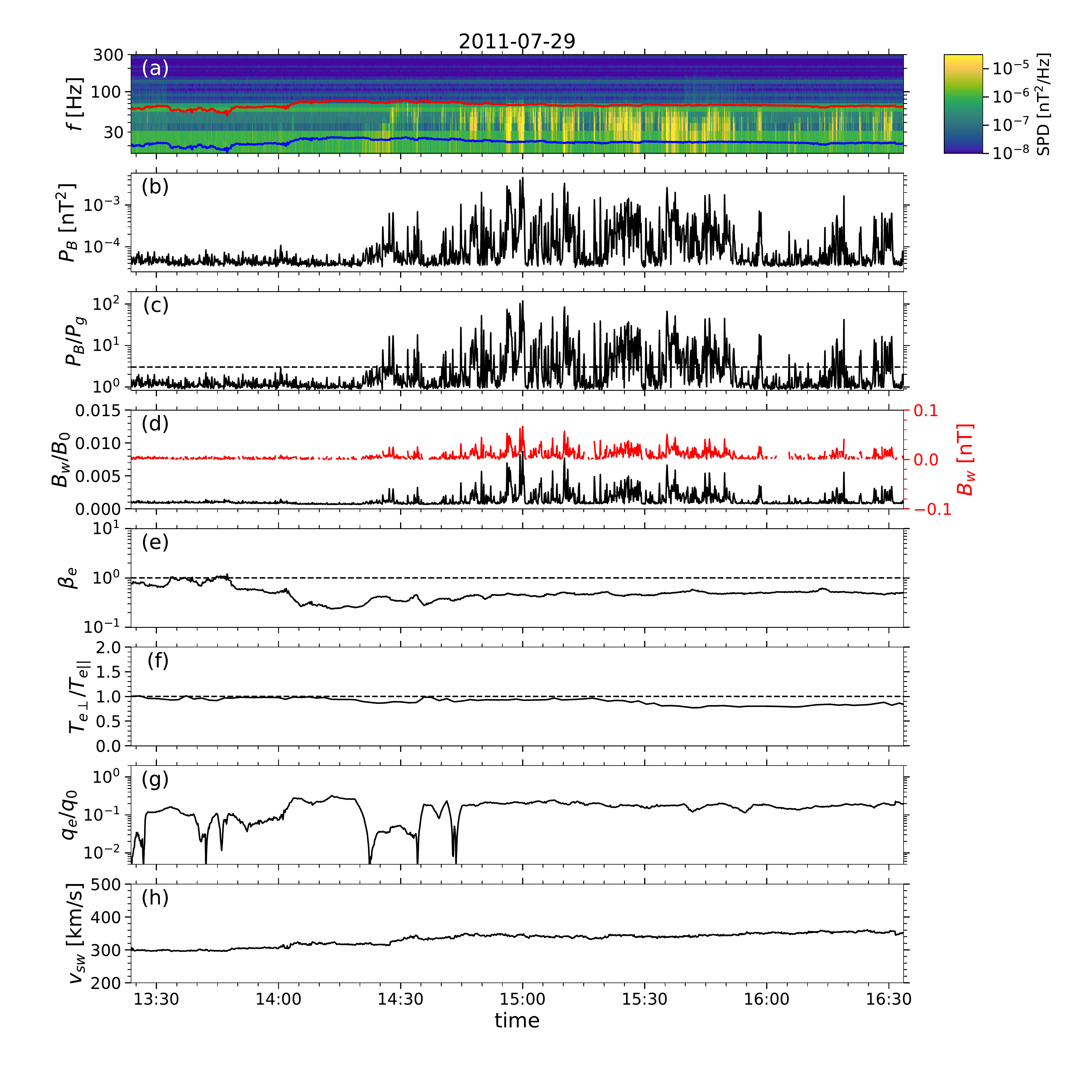}
    \caption{The wave activity in the whistler frequency range observed aboard ARTEMIS on July 29, 2011 (one day from our dataset): (a) magnetic field spectral power density,  $0.1\;f_{ce}$ and $0.3\;f_{ce}$ are indicated with green and red curves, where $f_{ce}$ is a local electron cyclotron frequency; (b) the magnetic field power $P_B$ in the frequency range between 16 and 300 Hz determined by Eq. (\ref{eq2}); (c) the magnetic field power $P_B$ normalized to the background turbulence power $P_g$ determined every 2 hours as 20th percentile of $P_B$; the visual inspection of our dataset showed that $P_B>3\;P_g$ (dashed line) is a reasonable criterion for selecting the wave activity events in the whistler frequency range and filtering out variations of the turbulence background; (d) the amplitude of magnetic field fluctuations evaluated as $B_{w}=(P_B-P_g)^{1/2}$ (red trace) and $B_w/B_0$ (black trace) that is the amplitude with respect to the local background magnetic field $B_0$; (e)-(h) $\beta_{e}=8\pi n_e T_{e||}/B_0^2$, electron temperature anisotropy $T_{e\perp}/T_{e||}$, the electron heat flux $q_{e}$ normalized to the free-streaming heat flux $q_0=1.5 n_e T_{e} (T_e/m_e)^{1/2}$, solar wind velocity $v_{sw}$.}
    \label{fig:example_day_overview}
\end{figure*}

Figure \ref{fig:example_day_overview} presents the magnetic field spectrum and particle moments for a particular day (July 29, 2011) in our dataset. Panel (a) shows the total spectral power density from 16 to 300 Hz. The SPD enhancements between 20 and 60 Hz appear first around 14:25 UT and continue intermittently thereafter before about \replaced{17:00 UT}{16:30 UT}. In terms of a local electron cyclotron frequency $f_{ce}$, the observed SPD enhancements are between $0.1$ and $0.3\;f_{ce}$ which is in the whistler frequency range. The wave activity can be characterized by the total magnetic field power in the frequency range between 16 and 300 Hz
\replaced{
\begin{equation}
    P_{B}\equiv \int_{16}^{300} \mathrm {SPD}(f)\;df \nonumber
    %\label{eq2}
\end{equation}
}{
\begin{equation}
    P_{B}\equiv \int_{16 \text{ Hz}}^{300 \text{ Hz}} \mathrm {SPD}(f)\;df
    \label{eq2}
\end{equation}
}
Panel (b) demonstrates that $P_B$ well traces the SPD enhancements. In the absence of clear wave activities, $P_B$ is a mixture of the inherent turbulence background and instrument noise between 16 and 300 Hz. We divide the magnetic field spectra into two-hour chunks and define the background power $P_{g}$ as the 20th percentile of $P_{B}$ within every chunk. Panel (c) presents $P_B/P_g$, demonstrating thereby that the wave activity corresponds to $P_B$ significantly exceeding $P_g$. The amplitude of magnetic field fluctuations associated with the wave activity is characterized by $B_{w}=(P_B-P_{g})^{1/2}$. Panel (d) shows that the amplitude of the magnetic field fluctuations reaches 0.05 nT, while $B_{w}/B_0$, that is the amplitude of the magnetic field fluctuations with respect to the background magnetic field, does not exceed 0.01. We emphasize that $B_{w}$ is the amplitude averaged over 8s, while the actual \added{peak} amplitude may be larger due to intermittent appearance of the magnetic field fluctuations over 8s. Panels (e) to (h) present a few plasma parameters: $\beta_{e} = 8\pi n_e T_{e||}/B_0^2$, $T_{e\perp}/T_{e||}$ is the electron temperature anisotropy, $q_e/q_0$ is the electron heat flux normalized to the free-streaming heat flux $q_0=1.5\;n_e T_e \;(2T_e/m_e)^{1/2}$, $v_{sw}$ is the solar wind \added{proton} velocity. \added{In the above parameters, $n_e$ is the electron density, $T_{e\perp}$ and $T_{e||}$ are the perpendicular and parallel electron temperature, $B_0$ is the magnitude of the quasi-static magnetic field. Note we have used a natural unit system in which temperature has the unit eV. The Boltzmann constant is dropped throughout the paper.}

Visual inspections of the magnetic field spectra from our dataset show that SPD enhancements in the whistler frequency range are always below 300 Hz. The wave power $P_B$ in the frequency range between 16 and 300 Hz is found to be a good indicator of the wave activity. The spectral power density in the first (8 Hz) frequency channel is excluded from $P_B$ computation, because it provides strong and noisy contribution to $P_B$, so that the wave activity at $f\geq 16$ Hz could not be identified in $P_B$. Another reason for excluding the first channel is that it is more likely to be contaminated by low-frequency magnetic field fluctuations different from whistler waves (see below). Visual inspections show that $P_B>3 P_g$ is a reasonable empirical criterion for selecting noticeable wave activities between 16 and 300 Hz and filtering out spectra corresponding to variations of the turbulence background. The criterion $P_B>3 P_g$ selects 17,050 magnetic field spectra that is about 38 hours and about 2\% of the original dataset.

Although the selected wave activities are in the whistler frequency range, they do not necessarily represent whistler waves (see Section \ref{sec1} for discussion). The routinely available ARTEMIS measurements include only two component of spectral power densities that are not sufficient to determine wave vectors and polarizations of the selected wave activity events. Nevertheless, these components, namely, spectral power densities SPD$_{\perp}$ and SPD$_{||}$, along with results of the previous observations enable us to filter out events contradicting the whistler wave interpretation and provide a basis to argue that the major part of the selected events are whistler waves. The technique relies on the previous analysis of the magnetic field spectral matrix (spectra and cross-spectra up to 400 Hz) measurements provided by Cluster \citep{Lacombe14} and the analysis of magnetic field waveforms (frequencies up to 64 Hz resolved) provided by ARTEMIS \citep{Stansby16,Tong2019a}, which both showed that whistler waves in the pristine solar wind propagate quasi-parallel to the background magnetic field ${\mathbf B}_0$. The observations of quasi-parallel whistler waves are consistent with theoretical predictions of potential instabilities operating in the solar wind \citep[][]{Gary1994a,Gary12}. Oblique whistler waves may be present in the solar wind, but they are predicted to be electrostatic and, hence, not identifiable in the magnetic field spectra \citep{Vasko2019a}.

\begin{figure}
    \centering
    \includegraphics[width=.8\linewidth]{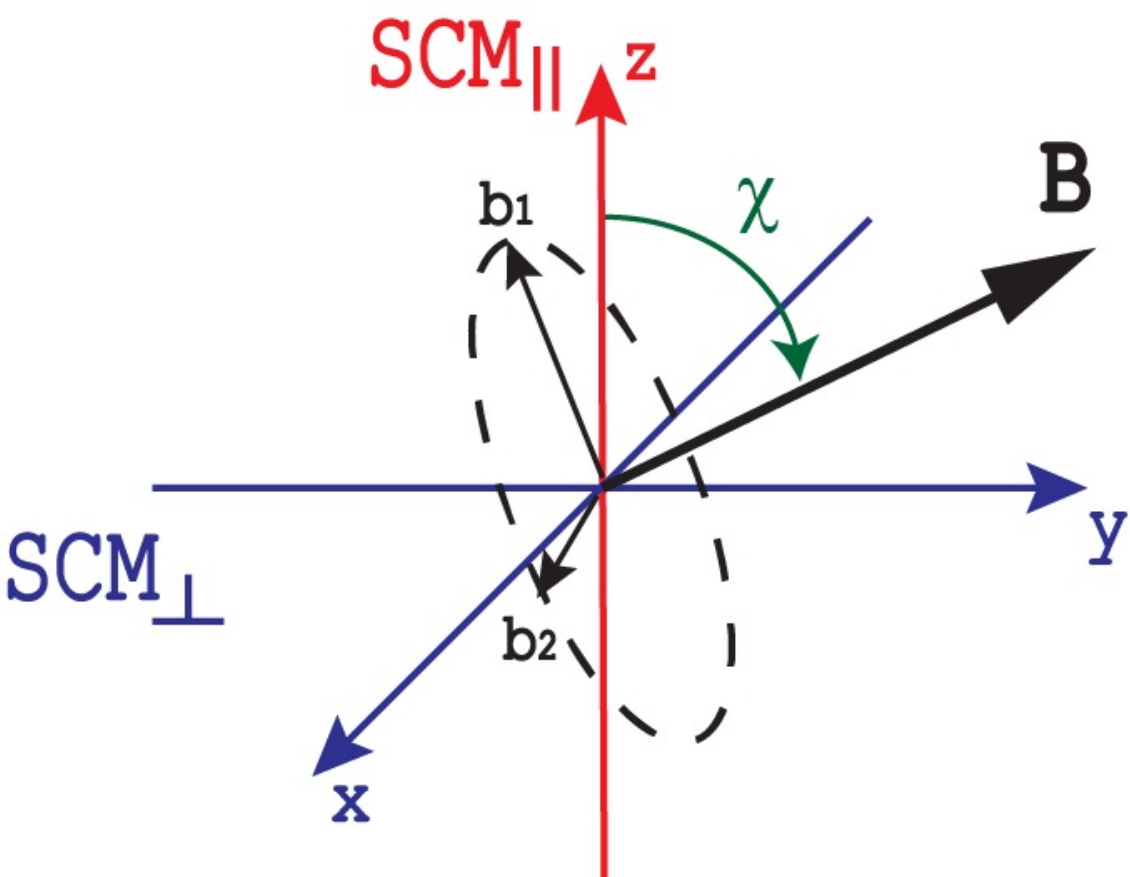}
    \caption{Schematics of the ARTEMIS search coil magnetometer antennas. The instrument provides spectral power densities SPD$_{||}$ and SPD$_{\perp}$ of magnetic field fluctuations along the spacecraft spin axis and in the plane perpendicular to the spin axis. The total spectral power density (Figure \ref{fig:example_day_overview}a) of the magnetic field fluctuations is computed as SPD=SPD$_{||}$+2$\;$SPD$_{\perp}$. For a whistler wave propagating parallel to the quasi-static magnetic field $B_0$ there is a particular relation between SPD$_{||}$ and SPD$_{\perp}$ that depends on angle $\chi$ (see Section \ref{sec2} for details).\label{fig2}}
\end{figure}

The whistler wave propagation parallel to the magnetic field results in a specific relation between SPD$_{||}$ and SPD$_{\perp}$ that is dependent on ${\mathbf B}_{0}$ orientation with respect to the spin axis (see Figure \ref{fig2} for schematics). A whistler wave at frequency $f$ propagating parallel to ${\mathbf B}_0$ is a circularly-polarized wave with the magnetic field along ${\mathbf b}_{1}\cos(2\pi f t)+{\mathbf b}_{2}\sin(2\pi f t)$, where ${\mathbf b}_{1,2}$ are unit vectors in the plane perpendicular to ${\mathbf B}_0$. This wave would produce SPD$_{||}(f)\propto \sin^2\chi$ and SPD$_{\perp}(f)\propto (1+\cos^2\chi)/2$, where $\chi$ is the angle between ${\mathbf B}_0$ and the spin axis (Fig. \ref{fig2}), so that the ratio
\begin{eqnarray}
R\equiv\frac{{\mathrm{SPD}}_{||}(f)}{{\mathrm{SPD}}(f)}
\label{eq:spd}
\end{eqnarray}
would equal to $R_{0}=0.5 \sin^2\chi$. A reasonable agreement between the observed $R$ and expected $R_0$ may allow filtering out events corresponding to plasma modes different from quasi-parallel whistler waves.

\begin{figure*}
    \centering
    \includegraphics[width=0.8\linewidth]{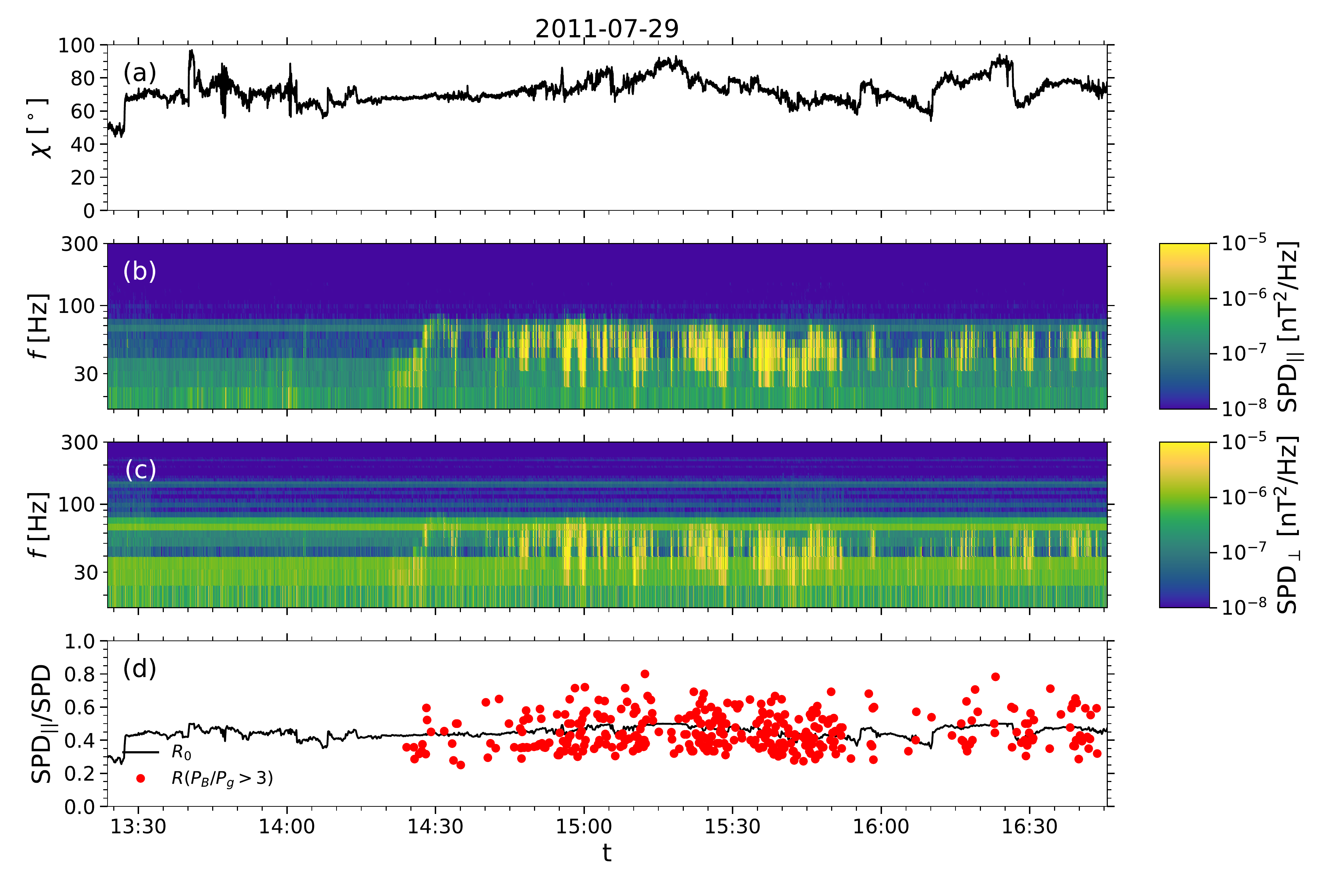}
    \caption{The test of the nature of the wave activity observed on July 29, 2011: (a) the angle $\chi$ between the magnetic field and the spin axis shown in Figure \ref{fig2} and computed using the quasi-static magnetic field measurements; (b, c) spectra SPD$_{||}$ and SPD$_{\perp}$ of magnetic field fluctuations along the spin axis and in the plane perpendicular to the spin axis; (d) the ratio $R={\mathrm{SPD}}_{||}(f_w)/{\mathrm{SPD}}(f_w)$ at the frequency channel $f_{w}$ corresponding to the largest SPD, only points at the moments of time with $P_{B}>3 P_{g}$ are indicated (red dots); the ratio $R_{0}$ expected for a whistler wave propagating parallel to the background magnetic field is shown by the black curve. \label{fig3}}
\end{figure*}

Figure \ref{fig3} presents the analysis of the nature of the wave activity shown in Figure \ref{fig:example_day_overview}. Panel (a) presents angle $\chi$ (Figure \ref{fig2}) computed using the quasi-static magnetic field measurements. Panels (b) and (c) present SPD$_{||}$ and SPD$_{\perp}$. For every magnetic field spectrum with $P_B>3P_g$ we identify the frequency channel $f_{w}$ with the largest total spectral power density, SPD in Figure \ref{fig:example_day_overview}a, and compute $R$ using SPD$_{||}$($f_w$) and SPD$(f_w)$ in Eq. (\ref{eq:spd}). Panel (d) shows that $R$ is well consistent with $R_0=0.5 \sin^2\chi$, supporting thereby the interpretation of the wave activity in terms of quasi-parallel whistler waves. 

\begin{figure*}
    \centering
    \includegraphics[width=1.0\linewidth]{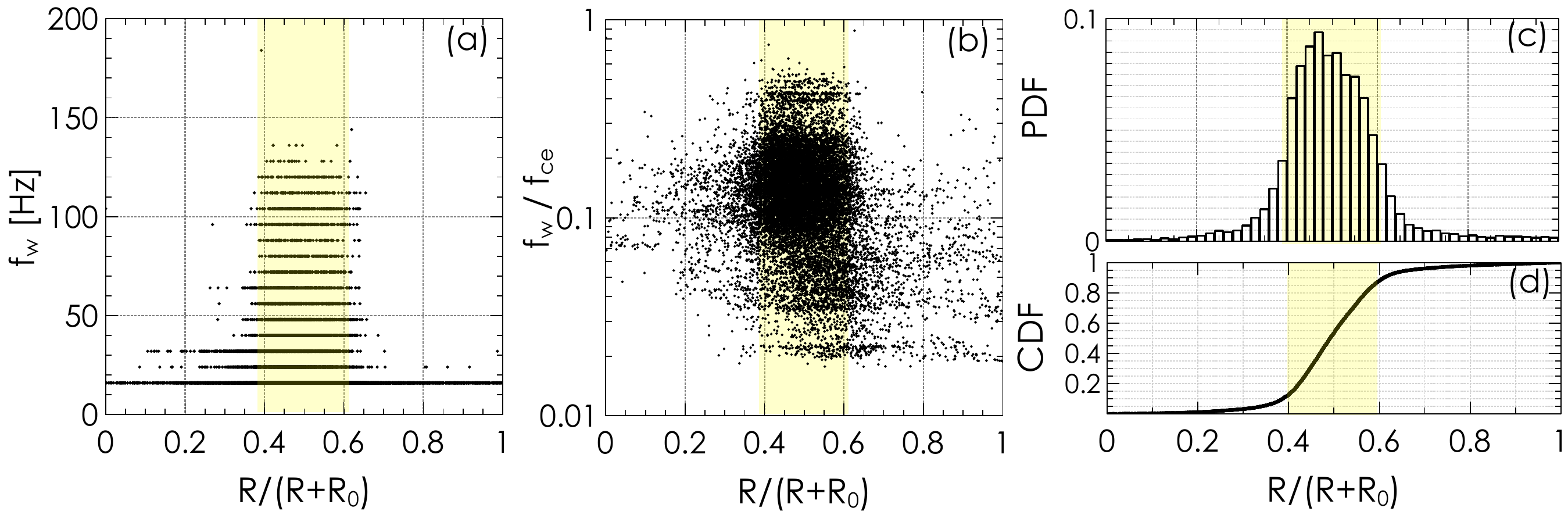}
    \caption{Results of testing the nature of the selected $\sim$17,050 magnetic field spectra through comparison of the observed $R={\mathrm{SPD}}_{||}(f_w)/{\mathrm{SPD}}(f_w)$, where $f_w$ is the frequency channel with the maximum SPD=SPD$_{||}$+2$\;$SPD$_{\perp}$, and $R$ value (denoted as $R_{0}$) expected for a whistler wave propagating parallel to the background magnetic field: (a,b) $R/(R+R_{0})$ vs. frequency $f_w$ and $f_w/f_{ce}$; (c,d) the probability and cumulative distribution functions of $R/(R+R_{0})$. The data points within the shaded region, $0.4<R/(R+R_0)<0.6$, correspond to wave activity events non-contradicting to the hypothesis of quasi-parallel whistler waves. Panel (d) shows that exclusion of the data points outside of the shaded region filters out less than \replaced{15\%}{20\%} of the data points.\label{fig4}}
\end{figure*}

Figure \ref{fig4} presents results of the comparison between $R$ and $R_0$ evaluated for all 17,050 magnetic field spectra with $P_B>3 P_g$. Panel (a) shows that $R/(R+R_0)$ are clustered around 0.5, that is $R\approx R_0$. Most of the events with $R/(R+R_0)$ significantly deviating from 0.5 are in the three lowest frequency channels at 16, 24 and 32 Hz, where low-frequency modes are expected most likely to appear due to the Doppler effect. Panel (b) shows that the events with $R/(R+R_0)$ significantly deviating from 0.5 have frequencies from 0.02 to 0.5 $f_{ce}$, demonstrating thereby that the whistler frequency range may be populated by plasma modes different from quasi-parallel whistler waves. We introduce a quantitative criterion $0.4<R/(R+R_0)<0.6$ to select the events not contradicting the interpretation of quasi-parallel whistler waves. The probability and cumulative distribution functions in panels (c) and (d) show that this selection criterion filters out about 20\% of the events leaving about 13,700 magnetic field spectra. In accordance with \cite{Lacombe14} this shows that whistler waves identified in the magnetic field spectra in the pristine solar wind are predominantly quasi-parallel. In what follows we use the selected 13,700 events to clarify how the occurrence and properties of whistler waves depend on macroscopic plasma parameters. 

\begin{figure}
    \centering
    \includegraphics[width=\linewidth]{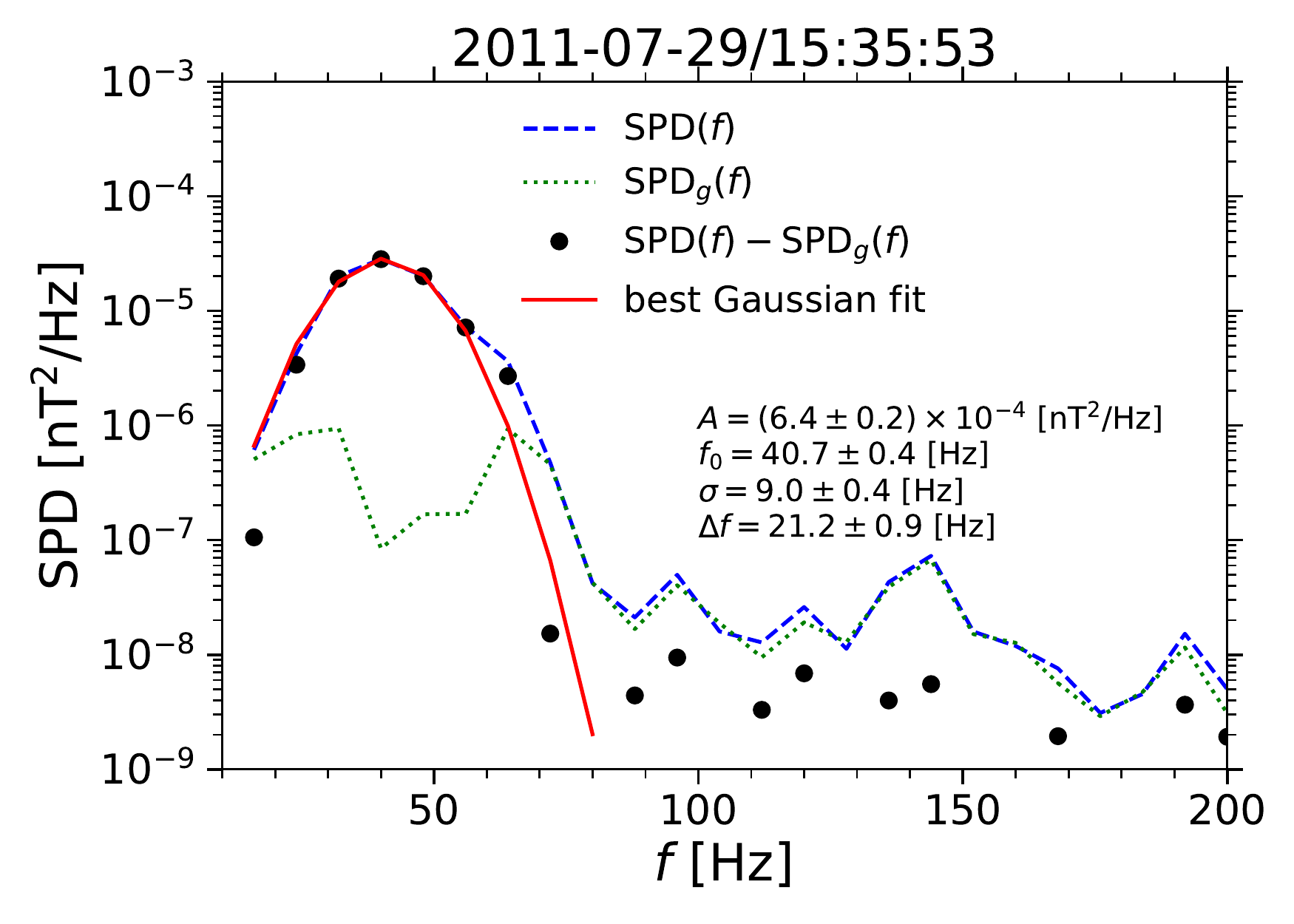}
    \caption{The analysis of the frequency bandwidth of a particular whistler wave spectrum on July 29, 2011. The measured spectral power density at 15:35:33 UT (blue), the background spectral power density SPD$_g(f)$ (green) that is computed at each frequency $f$ as the 20th percentile of SPD$(f)$ at that frequency every two hours. Black dots represent SPD$(f)$-SPD$_g(f)$ that is the whistler wave spectrum. The best-fit Gaussian (\ref{eq:Gauss}) to the whistler wave spectrum is given by the red curve. The best fit parameters $A$ and $\sigma$ are indicated along with the frequency bandwidth estimated as the width at half maximum, $\Delta f=2\sigma(2\ln 2)^{1/2}$.}
    \label{fig:example_guassian_fit}
\end{figure}

The selected whistler wave SPD enhancements spread over several frequency channels. To quantify the frequency bandwidth of the whistler waves, we determine first the background spectral power density SPD$_g(f)$ at frequency $f$ as the 20th percentile of SPD$(f)$ at that frequency every two hours. Similarly to $P_g$, SPD$_g(f)$ is a combination of the magnetic field turbulence background and intrinsic instrument noise level. The whistler wave spectrum SPD$(f)$-SPD$_g(f)$ is fitted to the Gaussian model with the peak at $f_{w}$
\begin{equation}
    \textrm{SPD}(f)-\textrm{SPD}_g(f)= A\exp\left[-\frac{(f-f_w)^2}{2\sigma^2}\right],
    \label{eq:Gauss}
\end{equation}
where $A$ and $\sigma$ are the best fit parameters. The frequency bandwidth $\Delta f$ is estimated as the full width at half maximum $$\Delta f = 2\sigma(2\ln 2)^{1/2}\sim 2.35\; \sigma$$ 

Figure \ref{fig:example_guassian_fit} presents the analysis of the frequency bandwidth of a particular whistler wave spectrum with the peak at $f_{w}\sim 40$ Hz measured at 15:35:33 UT on July 29, 2011 (one spectrum from Figure \ref{fig:example_day_overview}). The whistler wave SPD enhancement is about two orders of magnitude larger than SPD$_g(f_w)$. The  Gaussian fit to $\textrm{SPD}(f)-\textrm{SPD}_g(f)$ yields the frequency bandwidth $\Delta f\sim 21$ Hz. We restrict the statistical analysis of the frequency bandwidth to whistler wave events with $f_{w}>16$ Hz, because only in those events we could guarantee that the peak of the Gaussian is at $f_{w}$, rather than at some frequency below 16 Hz. The criterion $f_{w}>16$ Hz leaves 5,800 spectra for the frequency bandwidth analysis that is 42\% of the selected 13,700 whistler wave spectra.

\section{Whistler wave occurrence \label{sec3}}

Out of about $8\times 10^5$ spectra we have associated about 13,700 spectra with quasi-parallel whistler waves that yields a total occurrence probability of whistler waves of 1.7\%. We emphasize that this is the probability of sufficiently intense whistler waves ($P_B>3 P_g$) above 16 Hz, i.e. whistler waves that are less intense and at lower frequencies have been excluded. The overall occurrence of whistler waves in the pristine solar wind is certainly higher. We demonstrate below that the occurrence probability of the selected whistler waves depends on macroscopic plasma parameters.

\begin{figure*}[h]
    \centering
    \includegraphics[width=\linewidth]{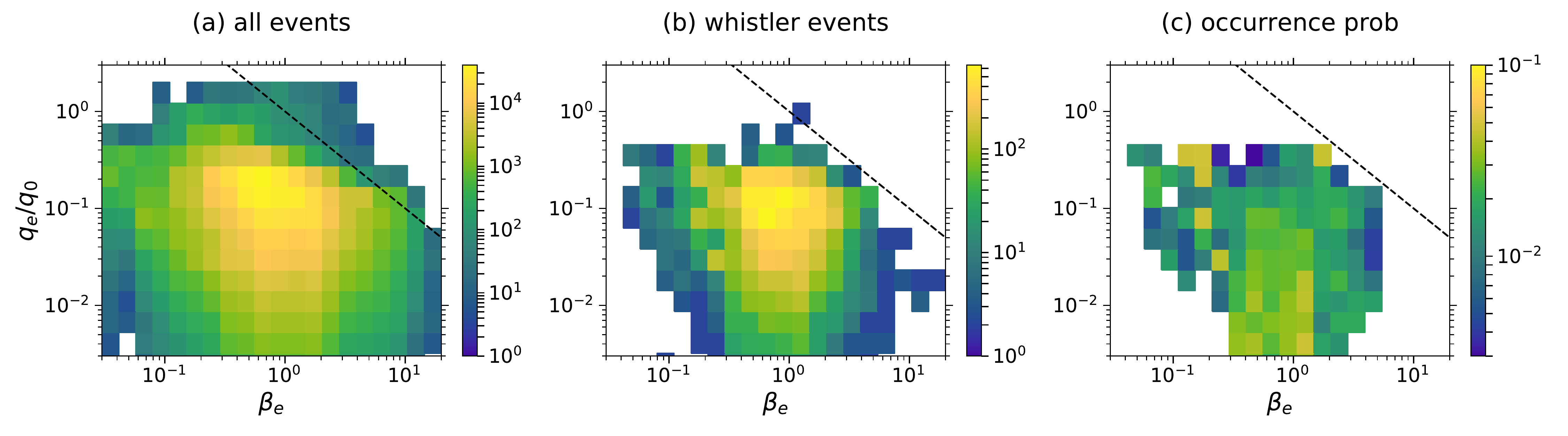}
    \caption{The analysis of whistler wave occurrence in dependence on the electron heat flux $q_e/q_0$ and $\beta_{e}$: (a) distribution of all $\sim8\times 10^5$ magnetic field spectra in $(q_e/q_0,\beta_{e})$ parameter plane; (b) distribution of the selected $\sim$13,700 spectra associated with quasi-parallel whistler waves; (c) the occurrence probability of whistler waves that is computed by dividing a number of events with whistler waves shown in panel (b) over a total number of events shown in panel (a). The dashed line in the panels represent $q_e/q_0=1/\beta_e$ for reference.}
    \label{fig:qeq0_vs_beta}
\end{figure*}

Figure \ref{fig:qeq0_vs_beta} presents the analysis of effects of the electron heat flux $q_{e}/q_0$ and $\beta_{e}$ on the occurrence probability of whistler waves. Panel (a) shows the distribution of all $\sim 8\times 10^5$ magnetic field spectra in $(q_e/q_0,\beta_{e})$ parameter plane. The electron heat flux at $\beta_{e}\gtrsim 1$ is below a threshold $q_e/q_0 \sim 1/\beta_{e}$ that is in agreement with previous spacecraft observations \citep{Gary1999b,Tong18}. This heat flux threshold was previously considered as the evidence for the heat flux regulation by the whistler heat flux instability \citep[][]{Feldman76,Gary1999b}. Panel (b) shows the distribution of $\sim 13,700$ magnetic field spectra associated with quasi-parallel whistler waves. Combining the distributions shown in panels (a) and (b) we evaluate the occurrence probability of whistler waves at various $(q_e/q_0,\beta_{e})$. Panel (c) shows that the occurrence probability does not favor the parameter space near the threshold $q_e/q_0 \sim 1/\beta_{e}$ and, instead, somewhat enhances at low heat flux values.

\begin{figure*}[h]
    \centering
    \includegraphics[width=\linewidth]{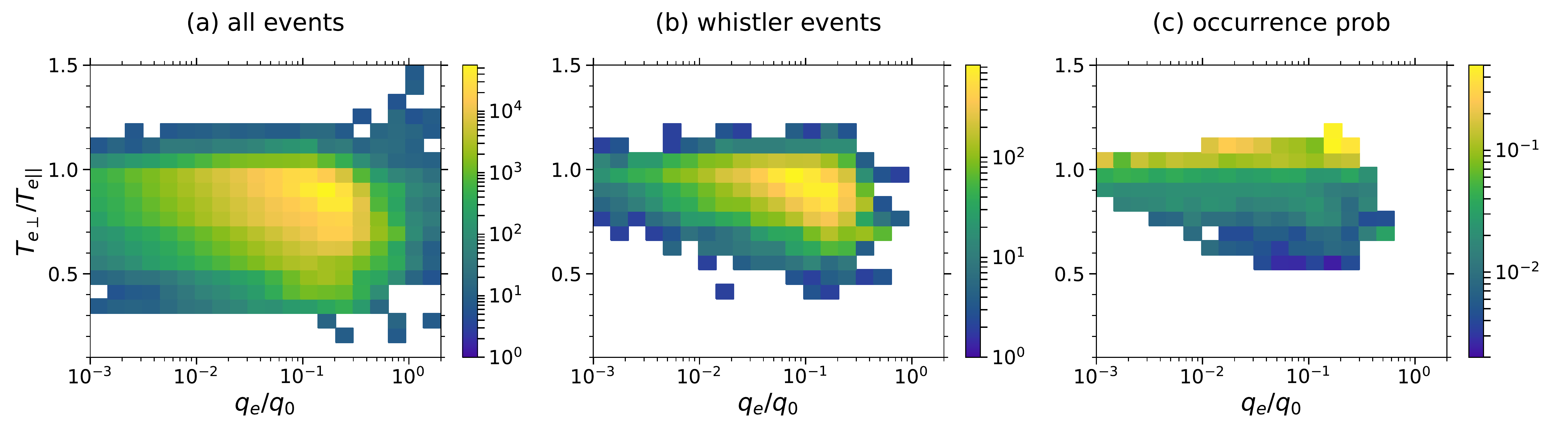}
    \caption{The analysis of whistler wave occurrence in dependence on the electron heat flux $q_e/q_0$ and $T_{e\perp}/T_{e||}$: (a) distribution of all $\sim8\times 10^5$ magnetic field spectra in $(q_e/q_0,T_{e\perp}/T_{e||})$ parameter plane; (b) distribution of the selected $\sim$13,700 spectra associated with quasi-parallel whistler waves; (c) the occurrence probability of whistler waves that is computed by dividing a number of events with whistler waves shown in panel (b) over a total number of events shown in panel (a).}
    \label{fig:anisotropy_vs_qeq0}
\end{figure*}

Figure \ref{fig:anisotropy_vs_qeq0} presents the analysis of effects of the electron heat flux $q_{e}/q_0$ and electron temperature anisotropy  $T_{e\perp}/T_{e||}$ on the whistler wave occurrence probability. Panels (a) and (b) present distributions of all events and whistler wave events in $(q_e/q_0,T_{e\perp}/T_{e||})$ parameter plane. In accordance with previous statistical studies \citep[e.g.,][]{Stverak08,Artemyev18:jgr} solar wind electrons at 1 AU most often exhibit parallel temperature anisotropy, $T_{e\perp}/T_{e||}<1$. Panels (a) and (b) are combined to compute the occurrence probability in $(q_{e}/q_0,T_{e\perp}/T_{e||})$ parameter plane. Panel (c) clearly demonstrates that the temperature anisotropy quite critically affects the whistler wave occurrence probability. At any given $q_{e}/q_0$ the occurrence probability increases with increasing $T_{e\perp}/T_{e||}$. The occurrence probability is less than a few percent at $T_{e\perp}/T_{e||}\lesssim 1$, but increases up to 10-60\% at $T_{e\perp}/T_{e||}>1$. In addition, \added{panel (b) shows} for whistler waves to occur the temperature anisotropy should be above a threshold that increases as the electron heat flux decreases: at $q_{e}/q_0\lesssim 10^{-2}$ the temperature anisotropy should be above 0.75, while at $q_{e}/q_0\gtrsim 3\times 10^{-2}$ whistler waves may occur at $T_{e\perp}/T_{e}||$ as low as 0.5. In addition to the 2D occurrence probabilities, we have computed whistler wave occurrence probabilities in dependence on individual macroscopic plasma parameters. 

\begin{figure*}
    \centering
    \includegraphics[width=\linewidth]{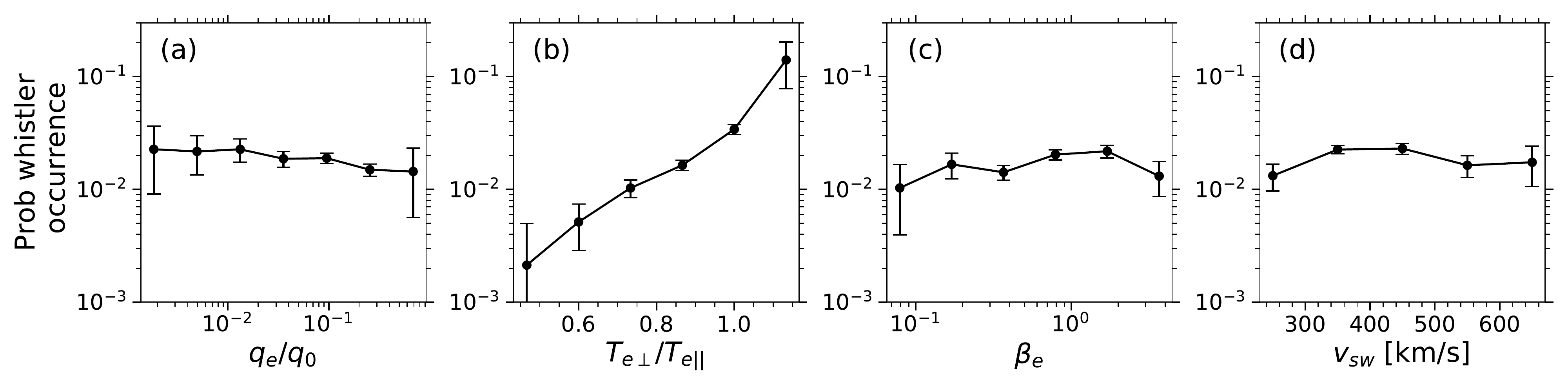}
    \caption{The occurrence probability of whistler waves in dependence on individual macroscopic plasma parameters.}
    \label{fig:occurrence_rate_1d}
\end{figure*}

Figure \ref{fig:occurrence_rate_1d} presents the occurrence probability of whistler waves in dependence on  $q_{e}/q_0$, $\beta_{e}$, $v_{sw}$ and $T_{e\perp}/T_{e||}$. The occurrence probability $P(\xi)$ of whistler waves in dependence on a macroscopic plasma parameter $A$ is determined as $P(\xi)=N_W(\xi)/N(\xi)$, where $N_W(\xi)$ is the number of whistler wave events with $A$ in the range $(\xi-\Delta \xi/2,\xi+\Delta \xi/2)$, while $N(\xi)$ is the total number of events with $A$ in the same range. The bin width $\Delta \xi$ is chosen so that the number of events within each bin would be sufficiently large. The uncertainties of $P(\xi)$ are estimated with the assumption \replaced{of independent Bernoulli trials}{that each particle measurement is independent \footnote{Assuming that each particle measurement has the same probability to have a whistler companion, and that $n$ measurements estimate the  probability to be $p$. Then the standard error of $p$ is $s_p = \sqrt{p(1-p)/n}$. We estimate the uncertainty of $p$ as the uncertainty at the 95\% level of confidence $\delta p = 2 s_p$.}}. Panels (a), (c) and (d) demonstrate that the electron heat flux, $\beta_e$ and solar wind velocity do not significantly affect the occurrence probability of whistler waves. Panel (b) confirms that the whistler wave occurrence probability is critically dependent on the electron temperature anisotropy. The probability is less than 2\% at $T_{e\perp}/T_{e||}<0.9$, but increases from 5 to 15\% as $T_{e\perp}/T_{e||}$ varies from 0.95 to 1.2.

\section{Whistler wave intensity \label{sec4}}

\begin{figure*}
    \centering
    \includegraphics[width=.8\linewidth]{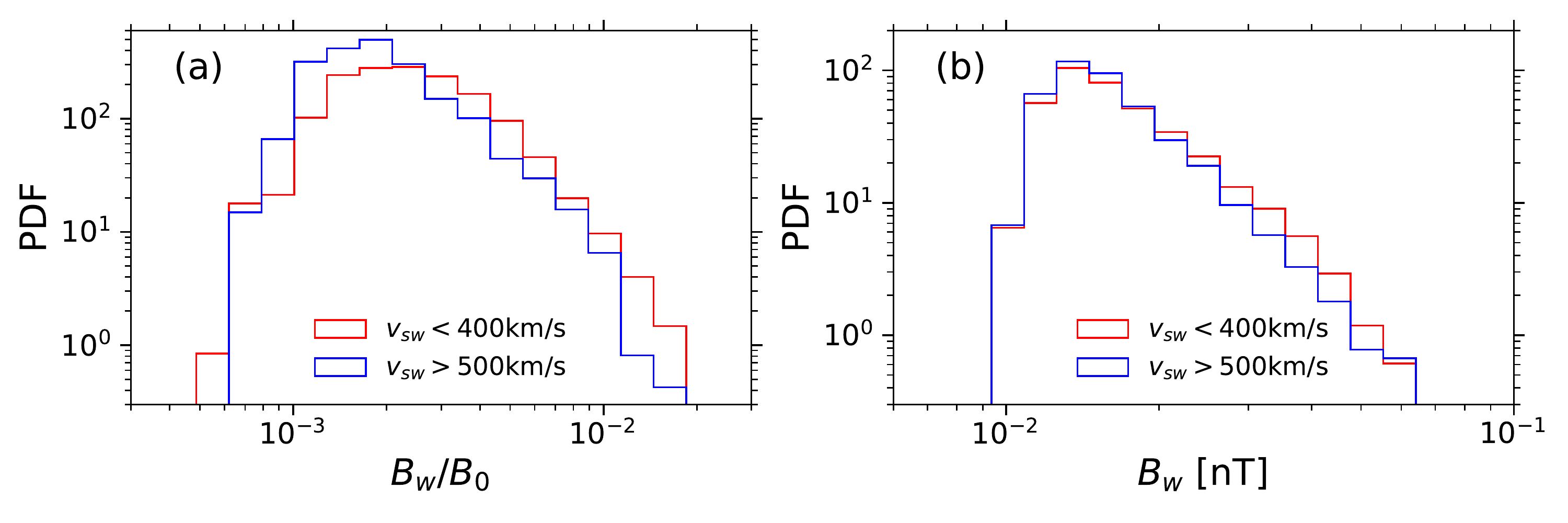}
    \caption{Probability distribution functions of whistler wave amplitudes $B_w$ and $B_w/B_0$ in the slow ($v_{sw}<400$ km/s) and fast ($v_{sw}>500$ km/s) solar wind.}
    \label{fig:db_vs_vsw}
\end{figure*}

Figure \ref{fig:db_vs_vsw} presents the probability distribution functions of whistler wave amplitudes $B_{w}$ and $B_{w}/B_0$ for the slow ($v_{sw}\lesssim 400$ km/s) and fast ($v_{sw}>500$ km/s) solar wind. Our dataset is dominated by the slow solar wind events, fast solar wind events constitute less than 12\% of the dataset. Panels (a) and (b) show that whistler waves amplitude $B_w$ is typically below 0.02 $B_0$ or in physical units in the range from 0.01 up to 0.1 nT. We recall that $B_w$ is the amplitude averaged over 8s, so that the actual \added{peak} amplitudes of magnetic field fluctuations could be in principle larger due to intermittent presence of whistler wave over 8s. However, these amplitudes are consistent with previous measurements of whistler waveforms aboard ARTEMIS spacecraft \citep{Stansby16,Tong2019a}, indicating thereby that quite likely whistler waves in the pristine solar wind have amplitudes $B_w$ much smaller than $B_0$. Panels (a) and (b) also demonstrate that there is a bit higher chance to observe intense whistler waves in the slow solar wind than in the fast solar wind.

\begin{figure*}
    \centering
    \includegraphics[width=\linewidth]{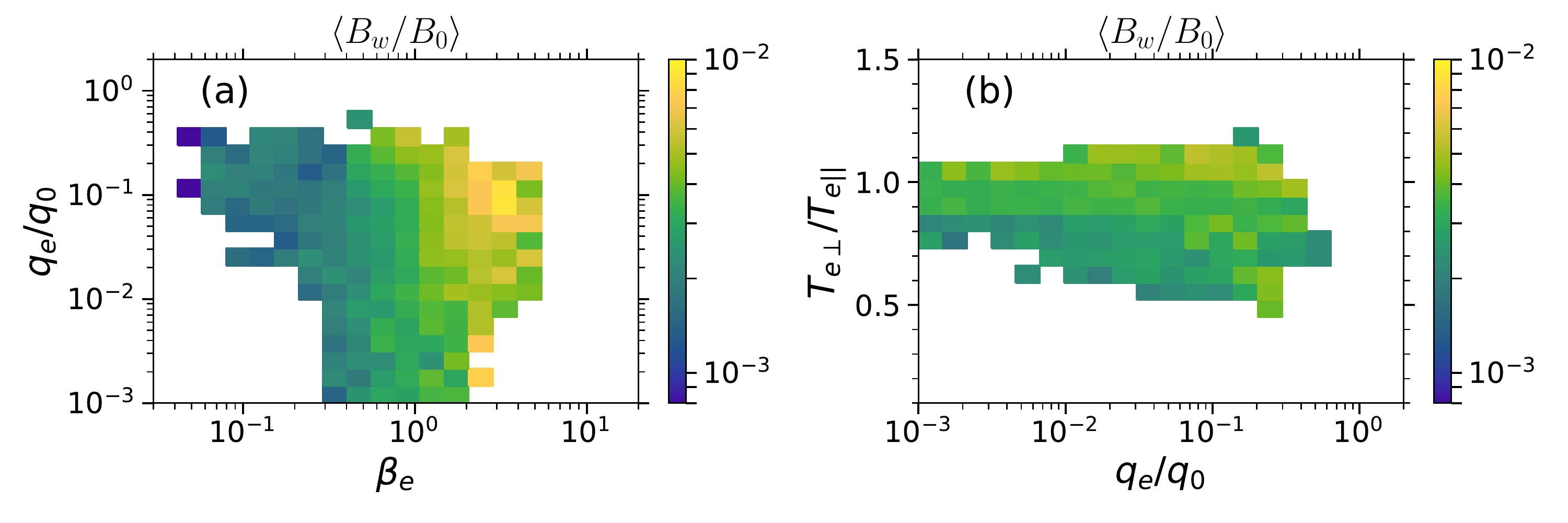}
    \caption{The whistler wave amplitude $\left<B_w/B_0\right>$ averaged over bins in  (a) $(q_e/q_0, \beta_e)$ and (b) $(T_{e\perp}/T_{e||}, q_e/q_0)$ parameter planes.} 
    \label{fig:amplitude_in_2d_plane}
\end{figure*}

Figure \ref{fig:amplitude_in_2d_plane} presents the distribution of the averaged whistler wave amplitude $\langle B_w/B_0\rangle$ in $(q_e/q_0,\beta_e)$ and $(T_{e\perp}/T_{e||},q_{e}/q_0)$ parameter planes. Panel (a) demonstrates that $\langle B_w/B_0\rangle$ is strongest, when both $\beta_{e}$ and $q_e/q_0$ are high. As a result, the averaged whistler wave amplitude is enhanced in the parameter space around to the threshold $q_{e}/q_0\sim 1/\beta_{e}$. It is interesting to note that the whistler wave occurrence probability doesn't favor this region in the parameter space (Figure \ref{fig:qeq0_vs_beta}). The reason is that the occurrence of whistler waves is most critically controlled by the temperature anisotropy, rather than $q_{e}/q_0$ or $\beta_{e}$. Panel (b) shows that $\langle B_{w}/B_0\rangle$
enhances with increasing $T_{e\perp}/T_{e||}$ at fixed $q_e/q_0$, while the positive correlation between $\langle B_w/B_0\rangle$ and $q_e/q_0$ is noticeable only at $T_{e\perp}/T_{e||}\gtrsim 1$.

\begin{figure*}
    \centering
    \includegraphics[width=\linewidth]{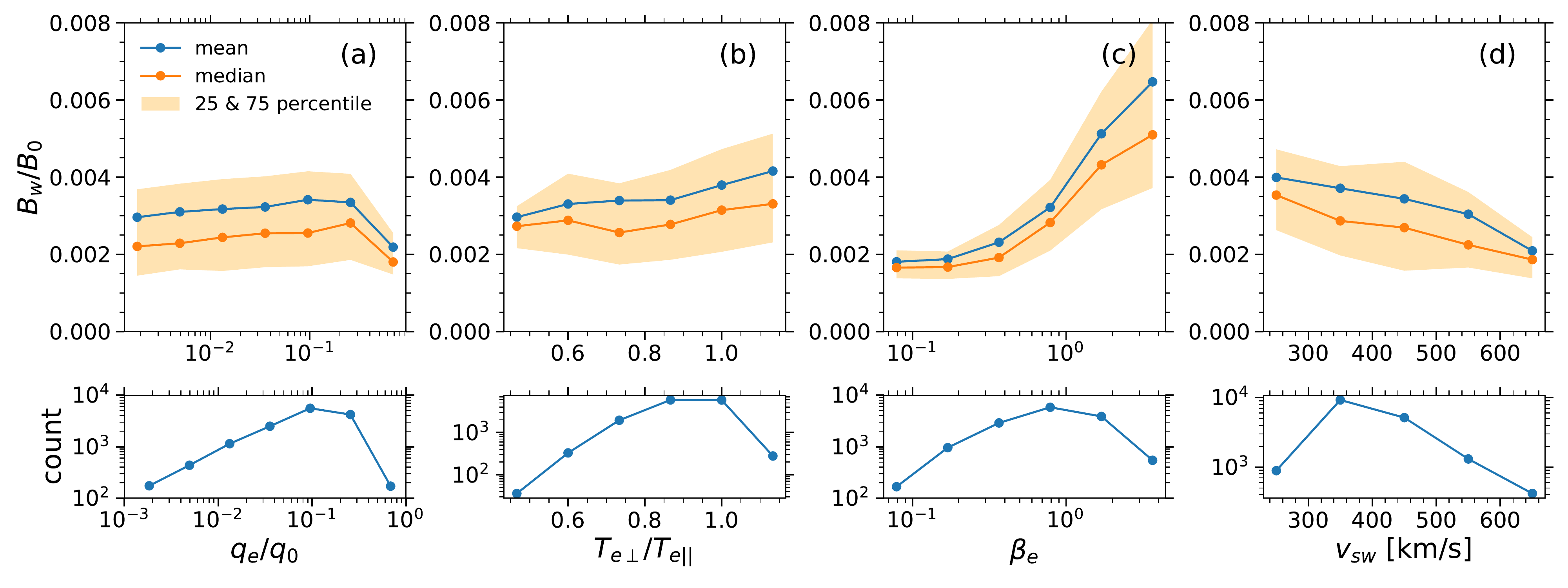}
    \caption{The whistler wave amplitude $B_w/B_0$ versus (a) the electron heat flux, (b) electron temperature anisotropy, (c) $\beta_{e}$, and (d) the solar wind velocity. The curves represent the median and mean values of $B_w/B_0$, while the shaded regions cover from 25th to 75th percentile of $B_w/B_0$.}
    \label{fig:wave_1d_panel}
\end{figure*}

\begin{figure}[]
    \centering
    \includegraphics[width=\linewidth]{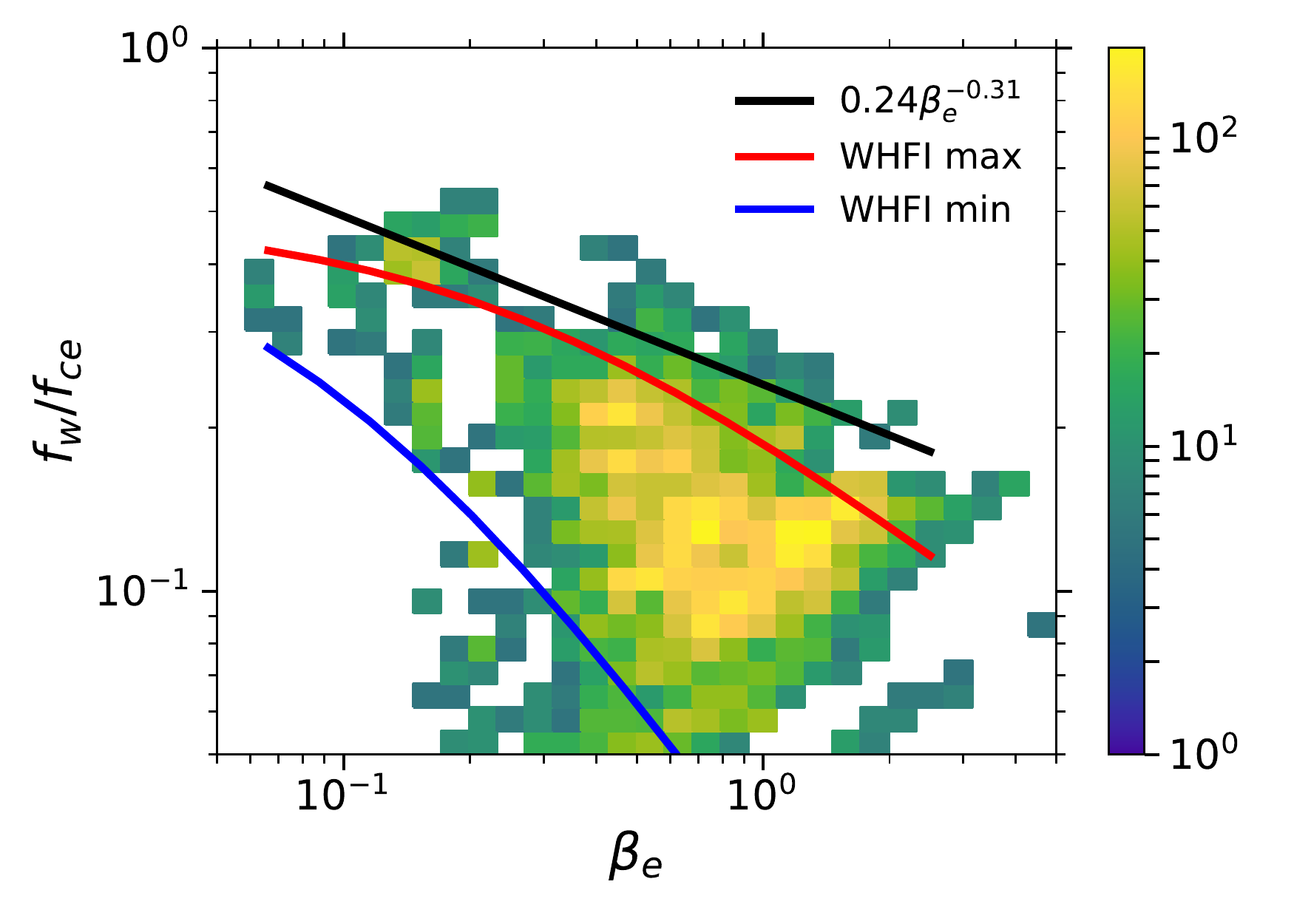}
    \caption{Whistler wave frequency $f_{w}$, determined as the frequency channel with the largest SPD$(f)$, normalized to the electron cyclotron frequency $f_{ce}$ versus $\beta_{e}$. \replaced{The orange dots represent 10\% of the highest frequency events at various $\beta_{e}$, the best power-law fit to these events is indicated by the red curve.}{The black curve represent the the best power-law fit to the 10\% of the highest frequency events at various $\beta_{e}$.} The \replaced{blue}{red} and \replaced{green}{blue} curves represent the maximum and minimum frequencies of whistler waves that can be generated by the whistler heat flux instability (see Section \ref{sec4.2} for details). The presented frequencies $f_{w}$ are measured in the spacecraft frame, but the estimates of the Doppler-shift have shown that these frequencies differ from the plasma frame frequencies by less than 30\% (see Section \ref{sec4.1} for details).}
    \label{fig:freq_vs_beta}
\end{figure}

Figure \ref{fig:wave_1d_panel} presents the distribution of whistler wave amplitudes $B_w/B_0$ in dependence on individual macroscopic parameters. The upper panels indicate the mean and median $B_w/B_0$ values in dependence on  $q_e/q_0$, $T_{e\perp}/T_{e||}$, $\beta_{e}$ and $v_{sw}$, while the shaded regions cover from the 25th percentile to the 75th percentile of $B_w/B_0$. The bottom panels present the number of events within bins used to compute the $B_w/B_0$ distributions in the upper panels. Panels (a) and (b) show that the mean and median values of $B_w/B_0$ are positively correlated with $q_{e}/q_0$ and $T_{e\perp}/T_{e||}$, though the overall variation of these values is \replaced{less than by a factor of two}{about 30\%}. The negative correlation between $B_w/B_0$ and the heat flux at $q_e/q_0\gtrsim 0.3$ is likely physical effect, because the number of events in the corresponding bins is sufficiently large. Panel (c) shows that the median and mean values of $B_w/B_0$ are most strongly correlated with $\beta_{e}$, both values increase by about a factor of \replaced{four}{three} as $\beta_{e}$ increases from 0.1 to 5. Panel (d) shows that the whistler wave amplitude is negatively correlated with the solar wind velocity, varying by a factor of two from the slow to fast solar wind.

\section{Whistler wave frequency\label{sec5}}

\subsection{Observations \label{sec4.1}}

We consider the frequency channel $f_{w}$ with the largest SPD$(f)$ or largest enhancement ${\mathrm{SPD}}(f) - {\mathrm{SPD}}_g(f)$ (both provide the same frequency channel) as the frequency of a whistler wave event. We could consider the frequency channel with the largest relative SPD enhancement, ${\mathrm{SPD}}(f)/{\mathrm{SPD}}_g(f)$, as the whistler wave frequency estimate. Because ${\mathrm{SPD}}_g(f)$ is a monotonically decreasing function of the frequency, this approach provides frequencies higher than $f_{w}$, but we have found that the difference is less than 50\%. We use $f_{w}$ as the whistler wave frequency estimate, while the use of the other frequency would not affect any of our conclusions. We have found that among various macroscopic plasma parameters only $\beta_{e}$ correlates strongly with the normalized frequency $f_w/f_{ce}$. 

Figure \ref{fig:freq_vs_beta} demonstrates that \deleted{$f_w/f_{ce}$ is negatively correlated with $\beta_{e}$. }\replaced{There}{there} are apparent upper and lower frequency bounds that decrease with increasing $\beta_{e}$. Below we compare these bounds to theoretical predictions of the whistler heat flux instability. To quantify the negative correlation between \added{the upper bound on } $f_{w}/f_{ce}$ and $\beta_{e}$ we bin all the whistler wave events according to $\beta_{e}$ and select 10\% of the highest frequency events within each bin. These highest frequency events\deleted{, indicated in Figure \ref{fig:freq_vs_beta},} are fitted to a power-law of $\beta_{e}$. The best fit \added{(black curve)} shown in Figure \ref{fig:freq_vs_beta} demonstrates that we generally have $f_{w}/f_{ce}\lesssim 0.24\;\beta_{e}^{-0.31}$. The whistler wave frequencies in Figure \ref{fig:freq_vs_beta} are measured in the spacecraft frame and differ from those in the plasma frame by the Doppler-shift, $\Delta f_{D}={\mathbf k} {\mathbf v}_{sw}/2\pi$, where ${\mathbf k}$ is the whistler wave vector. We have estimated the Doppler-shift for all whistler waves events using the wave vector estimate from the cold dispersion relation, $f/f_{ce}=k^2d_e^2/(1+k^2d_e^2)$, where $d_{e}=c/\omega_{pe}$ is the electron inertial length and $\omega_{pe}$ is the electron plasma frequency \citep[e.g.,][]{Stix62}. We have found that $\Delta f_{D}/f_{w}$ is less than 0.3, so that the measured frequency can be considered as a good estimate of the whistler wave frequency in the plasma frame.

\begin{figure*}
    \centering
    \includegraphics[width=.8\linewidth]{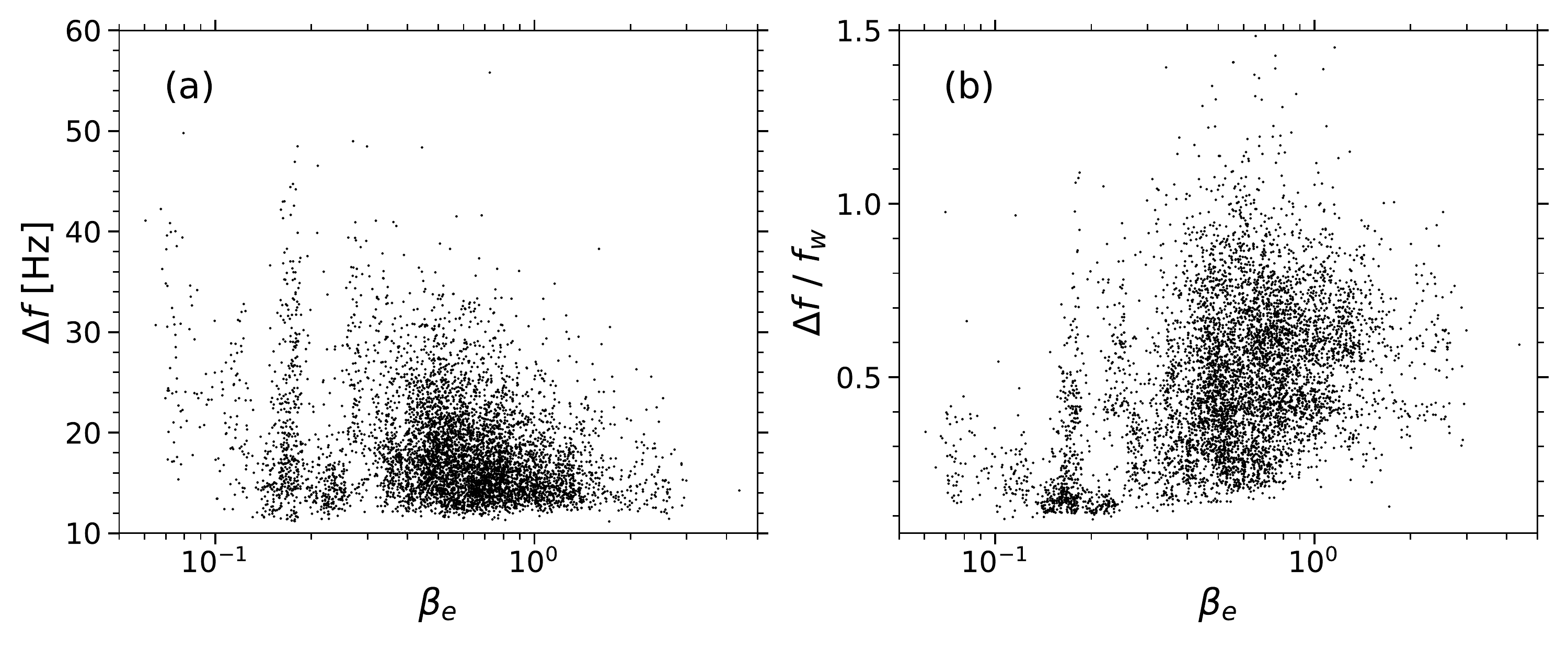}
    \caption{The frequency bandwidth, in physical units and normalized to $f_{w}$, of 5,800 whistler wave events, whose frequency $f_{w}$ is above 16 Hz. The frequency bandwidth is presented versus $\beta_{e}$.}
    \label{fig:df_f0_vs_beta}
\end{figure*}

Figure \ref{fig:df_f0_vs_beta} presents the frequency bandwidth $\Delta f$ of about 5,800 whistler wave events with $f_{w}>16$ Hz. Panel (a) shows that $\Delta f$ is typically about 15 Hz, though can be as large as 50 Hz\deleted{, and correlates negatively with $\beta_{e}$}. Panel (b) shows that the frequency bandwidth normalized to the whistler wave frequency $f_{w}$ is typically in the range between 0.1 and 1. There is a clear positive correlation between $\Delta f/f_{w}$ and $\beta_{e}$: at $\beta_{e}\ll 1$ whistler waves typically exhibit $\Delta f/f_{w}\sim 0.2$, while $\Delta f/f_{w}$ is typically about 0.5 at $\beta_{e}\sim 1$. The implications of the frequency width estimates will be discussed in Section \ref{sec6}.

\subsection{WHFI predictions \label{sec4.2}}

The linear theory of the WHFI suggests that the electron velocity distribution function (VDF) consisting of bi-Maxwellian core and halo populations, counter-streaming in the plasma rest frame, can be unstable to whistler wave generation at sufficiently large core and halo bulk velocities \citep{Gary75,Gary1994a}. \cite{Tong2019a} have recently shown for several events that the WHFI indeed generates whistler waves in the pristine solar wind. In this section we evaluate the maximum and minimum frequencies of whistler waves expected to be produced by the WHFI  in dependence on $\beta_{e}$. We consider the simplest electron VDFs consisting of isotropic core and halo populations ($T_{\perp}=T_{||}$) and assume a zero net electron current in the plasma rest frame, $n_{c}\Delta v_{c}+n_h\Delta v_{h}=0$, where $n_{c,h}$ and $\Delta v_{c,h}$ are densities and bulk velocities of the core and halo populations. Because the bulk velocities are much smaller than the corresponding thermal velocities \citep[e.g.,][]{Feldman75,Tong2019a}, we have $\beta_{e}\approx \beta_c+\beta_h$, where $\beta_{c}=8\pi n_{c} T_{c}/B_0^2$, $\beta_{h}=8\pi n_{h} T_{h}/B_0^2$ and $T_{c,h}$ are core and halo temperatures.

\begin{table}[]
    \centering
    \begin{tabular}{cc}
    \hline
    variable & values \\
    \hline
    $T_c/T_p$ & 1 \\
    $n_c/n_0$ & $\{0.75, 0.8, 0.9, 0.95\}$ \\
    $T_h/T_c$ & $\{3, 5, 7, 9, 11\}$ \\
    $\Delta v_c/v_A$ & $\{-i/2 | i=0,1,2....20\}$ \\
    \hline
    \end{tabular}
    \caption{Parameter ranges used for the analysis of the maximum and minimum frequencies of whistler waves that can be generated by the whistler heat flux instability (see Section \ref{sec4.2} for details).}
    \label{tab:WHFI_parameters}
\end{table}

The linear growth rate of the WHFI normalized to $f_{ce}$ depends on $n_c/n_0$, $T_h/T_c$, $T_p/T_c$,  and $\Delta v_c/v_A$, where $n_0$ is the total electron density \added{which is also assumed equal to the proton density}, $T_{c,h}$ are the core and the halo temperatures, $T_p$ is the proton temperature, and $v_A=B_0/(4\pi n_0 m_p)^{1/2}$ is the Alfv{\'e}n velocity \added{, and $m_p$ is the proton mass}. The growth rate is almost independent of the proton to core electron temperature ratio, because in realistic conditions protons do not resonate with whistler waves produced by the WHFI\added{ \citep{Gary75}}. In what follows we keep $T_p/T_c=1$ which is a reasonable assumption at 1 AU \citep[e.g.,][]{Newbury98,Artemyev18:jgr}. To evaluate the maximum and minimum frequencies of whistler waves that can be generated by the WHFI instability, we fix $\beta_{e}$ and vary $n_c/n_0$, $T_h/T_c$ and $\Delta v_c/v_A$ in the ranges typical for the solar wind at 1 AU (Table \ref{tab:WHFI_parameters}). For each combination of these three parameters we compute the linear growth rate using the numerical code developed by \cite{Tong15} and identify the frequency of the fastest growing whistler wave. Then, for each fixed $\beta_{e}$ we identify the maximum and minimum frequencies of whistler waves that can be generated by the WHFI. At a fixed $\beta_{e}$ the minimum frequency decreases with decreasing threshold value on the growth rate. Different threshold values result in different minimum frequency bounds, but these bounds are of similar shape and almost parallel to each other in the $(\beta_e, f/f_{ce})$ plane. The maximum and minimum frequency bounds are well fitted to modified power-laws
\begin{equation}
   f/f_{ce} = a (\beta_e + b)^c
   \label{eq:freq_power_law}
\end{equation}
Table \ref{tab:WHFI_freq_fits} presents the best fit parameters $a$, $b$ and $c$ for the maximum frequency bound\added{ at zero growth rate} and for the minimum frequency bounds derived for several growth rate thresholds, $\gamma/\omega_{ce}>10^{-5}$ and $10^{-6}$, where $\omega_{ce}=2\pi f_{ce}$.

\begin{table}[]
    \centering
    \begin{tabular}{ccccc}
    \hline
        & $\gamma/\omega_{ce}$ & $a$ & $b$ & $c$ \\
    \hline
        $f_{\max}/f_{ce}$ & $>0$ & 0.19 & 0.22 & -0.58  \\
        $f_{\min}/f_{ce}$ & $> 10^{-5}$ & 0.046 & 0.058 & -0.95\\
        $f_{\min}/f_{ce}$ & $> 10^{-6}$ & 0.034 & 0.09 & -1.1 \\
    \hline
    \end{tabular}
    \caption{Values of parameters $a, b$ and $c$ in Eq. (\ref{eq:freq_power_law}) that gives fitting to the maximum and minimum frequencies of whistler waves that can be generated by the whistler heat flux instability at various $\beta_{e}$. \added{The maximum frequency quickly converges to some asymptotic value as the growth rate tends to zero, whereas }\replaced{The}{the} minimum frequency bound depends on the growth rate threshold. We present parameters for \added{the maximum frequency bound at zero growth rate, and} the minimum frequency bounds computed for $\gamma/\omega_{ce}>10^{-5}$ and $10^{-6}$, where $\omega_{ce}=2\pi f_{ce}$.}
    \label{tab:WHFI_freq_fits}
\end{table}

Figure \ref{fig:freq_vs_beta} overlays the theoretical maximum and minimum frequency bounds upon the measured whistler wave frequencies. The presented minimum frequency bound is derived for $\gamma/\omega_{ce}>10^{-6}$. The frequencies of the major part of the observed whistler waves fall between the minimum and the maximum theoretical bounds, demonstrating thereby that the observed whistler waves could be in principle generated by the WHFI. Moreover, the generation can be local that is the whistler waves are generated in a local plasma, rather than generated in some other region and propagated to the spacecraft location.

\section{Discussion \label{sec6}}

We have carried out statistical analysis of whistler waves observed in the pristine solar wind using the most representative dataset collected up to date. We have focused on whistler waves identified by a local peak in the spectral power density of the magnetic field fluctuations, that is why these whistler waves are produced by free energy in a plasma, rather than by the turbulence cascade. Out of 801,527 magnetic field spectra measured at 1 AU aboard ARTEMIS, we have selected about 17,050 intense wave activity events in the whistler frequency range and associated 13,700 of them with quasi-parallel whistler waves. Thus, about 80\% of the intense events in the whistler frequency range are consistent with quasi-parallel whistler wave interpretation. This conclusion is in agreement with results of the previous less extensive studies of waveform and cross-spectra measurements \citep{Lacombe14,Stansby16,Tong2019a}. The other $\sim$20\% of the intense events are highly likely low-frequency plasma modes Doppler-shifted into the whistler frequency range, because they are predominantly observed in the three lowest frequency channels. The overall occurrence of quasi-parallel whistler waves in our dataset is about 1.7\%, but the actual occurrence of whistler waves is certainly higher, because we selected only sufficiently intense whistler waves above 16 Hz. 

We have shown that the occurrence probability of whistler waves most critically depends on the electron temperature anisotropy. There is no any drastic dependence of the whistler wave occurrence on the electron heat flux, solar wind velocity or $\beta_{e}$. The occurrence probability is less than 2\% when $T_{e\perp}/T_{e||}\lesssim 0.9$, but varies from 5 to 15\% as $T_{e\perp}/T_{e||}$ increases from 0.95 to 1.2. This correlation is consistent with the recent analysis by \cite{Tong2019a} of several whistler wave events measured in the burst mode (waveform available) aboard ARTEMIS. \cite{Tong2019a} have shown that whistler waves in those events were generated locally by the WHFI, while the temperature anisotropy of the halo population $T_{h\perp}/T_{h||}$ critically affects the instability onset: $T_{h\perp}/T_{h||}$ sufficiently smaller than unity quenches the instability, while $T_{h\perp}/T_{h||} > 1$ significantly enhances the growth rate. In the present statistical analysis we did not compute temperature anisotropies of the core and halo electron populations, but we expect that the increase of the full anisotropy $T_{e\perp}/T_{e||}$ corresponds to the increase of the halo temperature anisotropy, because temperature anisotropies of core and halo populations are positively correlated \citep[][]{Feldman76,Pierrard16}.

We have shown that whistler waves in the solar wind have amplitudes $B_w$ typically below 0.02 $B_0$ or in physical units below 0.1 nT. These amplitude estimates are consistent with the previous less extensive studies, where waveform measurements were analyzed \citep{Lacombe14,Stansby16,Tong2019a}, but more extensive waveform analysis should be carried out in the future to verify this result. The averaged whistler wave amplitude $B_w/B_0$ is found to be negatively correlated with the solar wind velocity. The average $B_w/B_0$ correlates positively with the electron heat flux and electron temperature anisotropy, but the strongest positive correlation is found with $\beta_{e}$. The variation of $q_{e}/q_0$ and $T_{e\perp}/T_{e||}$ over the observed range results in variation of $B_w/B_0$ by \replaced{a factor of two}{about 30\%}, while the variation of $\beta_{e}$ from 0.1 to 5 results in variation of $B_w/B_0$ by a factor of \replaced{four}{three}. The presented amplitude estimates and correlations between $B_w/B_0$ and macroscopic parameters should be useful for future theoretical studies of origin and effects of whistler waves in the solar wind. At the moment, we note that the whistler wave amplitudes observed at 1 AU are much smaller than whistler wave amplitudes $B_w\sim B_0$ reported in recent Particle-In-Cell simulations \citep{Roberg-Clark:2016,Roberg-Clark:2018b}, indicating thereby that the simulations are initialized with electron VDFs unrealistic for the solar wind at 1 AU. The fact that the whistler wave amplitudes are rather small calls into question their role in the electron heat flux regulation in the solar wind, though this question deserves a separate study.

\begin{figure*}
    \centering
    \includegraphics[width=0.8\linewidth]{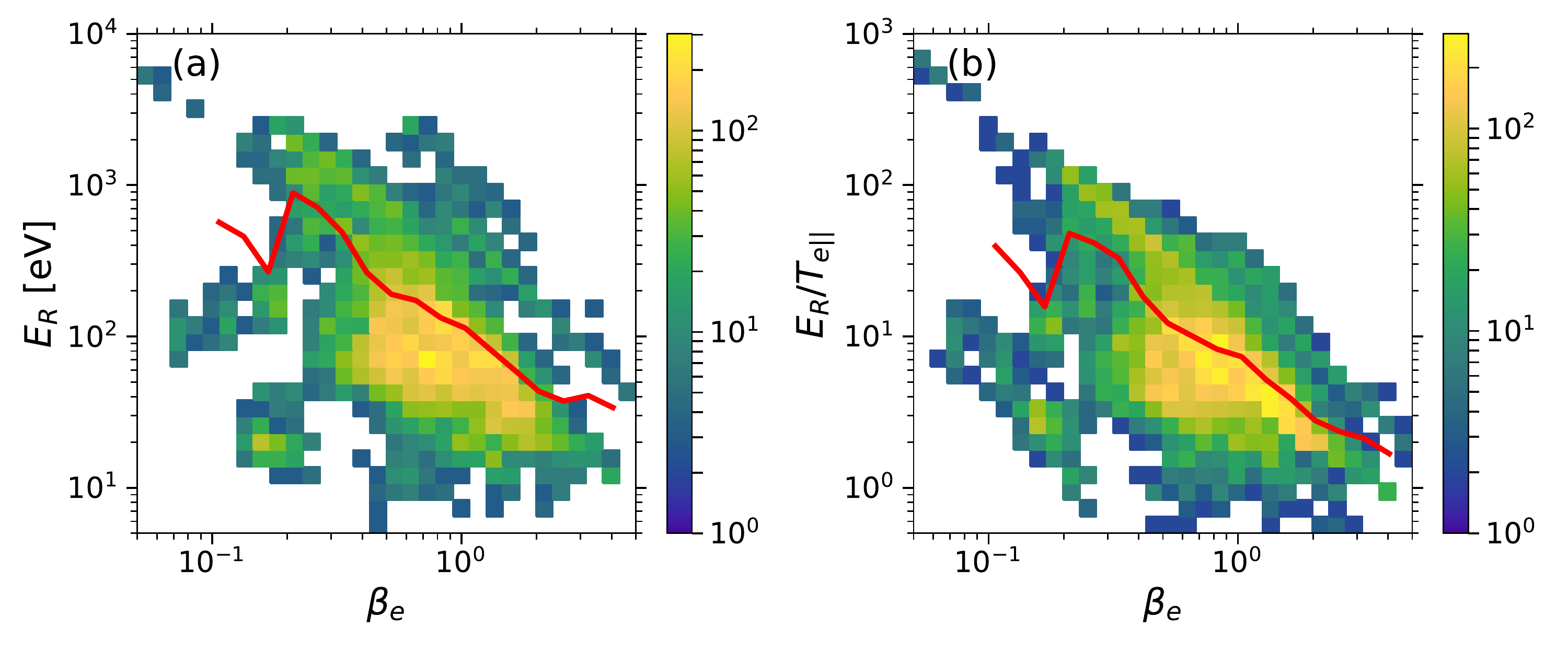}
    \caption{The minimum energy of electrons to be in the first normal cyclotron resonance with the observed whistler waves. It is given by Eq. (\ref{eq:Er}) with the whistler wave frequencies adopted from Figure \ref{fig:freq_vs_beta}a. Panel (a) presents the minimum resonant energy in physical units, while panel (b) presents this energy with respect to the electron temperature $T_{e||}$. The averaged resonant energies are presented by the red curves.}
    \label{fig12}
\end{figure*}

We have estimated the frequencies of the observed whistler waves and bandwidths of the whistler wave spectra. The only electrons that can drive and efficiently interact with quasi-parallel whistler waves are those in the first normal cyclotron resonance \citep[e.g.,][]{Shklyar09}
\begin{equation}
v_{||}=\frac{\omega-\omega_{ce}}{k},
\end{equation}
where $v_{||}$ is electron velocity parallel to the quasi-static magnetic field, $\omega=2\pi f$, $\omega_{ce}=2\pi f_{ce}$ and $k$ is the whistler wavenumber. The minimum energy of the cyclotron resonant electrons \citep[e.g.,][]{Kennel66}

\begin{equation}
E_{R}=\frac{B_0^2}{8\pi n_0}\frac{f_{ce}}{f}\left(1-\frac{f}{f_{ce}}\right)^3,
\label{eq:Er}
\end{equation}
where we have used cold dispersion relation of whistler waves, $f/f_{ce}=k^2d_e^2/(1+k^2d_e^2)$ \citep[e.g,][]{Stix62}. Figure \ref{fig12} presents the minimum resonant energy evaluated using Eq. (\ref{eq:Er}) with whistler wave frequencies adopted from Figure \ref{fig:freq_vs_beta}a. The minimum resonant energy is negatively correlated with $\beta_{e}$, because $E_R\propto B_0^2$, while $\beta_{e}\propto 1/B_0^2$. Panel (a) shows that the minimum resonant energy is of a few tens of eV at $\beta_{e}\sim 1$ and above a few hundred eV at low $\beta_{e}$. Panel (b) shows that in terms of thermal energies the resonant energy is about $3\;T_{e}$ at $\beta_{e}\sim 1$ and a few tens of $T_{e}$ at low $\beta_{e}$. We conclude that the observed quasi-parallel whistler waves should be driven by the halo electron population in accordance with previous theoretical \citep[][]{Gary75,Gary1994a} and experimental \citep[][]{Tong2019a} studies.

\begin{figure}
    \centering
    \includegraphics[width=\linewidth]{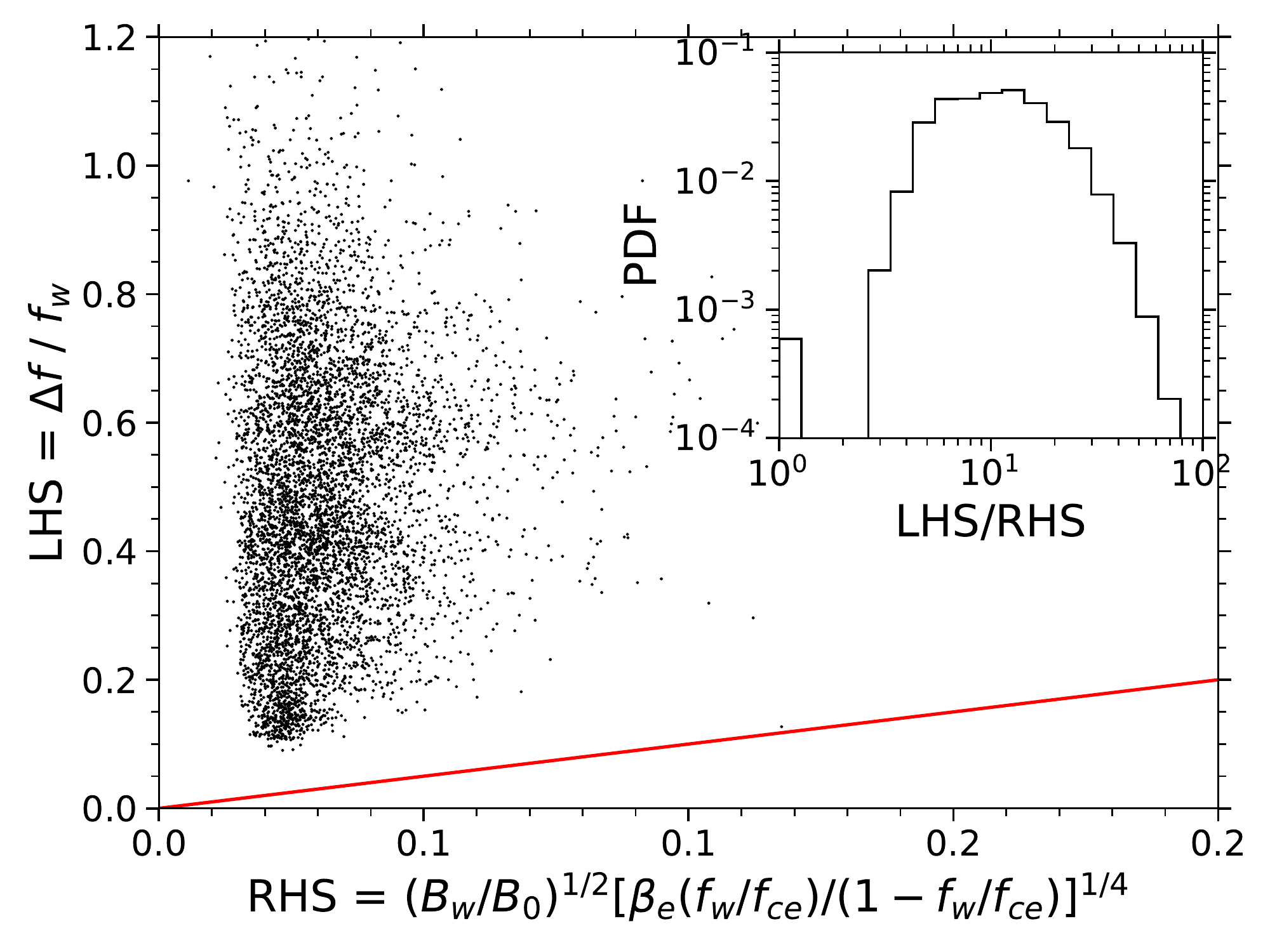}
    \caption{Estimated values of the left hand side (LHS) and the right hand side (RHS) of Eq. (\ref{eq:qlt_appl}) using ARTEMIS measurements. The red line references equality between LHS and RHS. The probability density function of the ratio LHS/RHS is shown in the inset panel.}
    \label{fig:QLT_applicability}
\end{figure}

The estimated bandwidths of the whistler wave spectra allow us to evaluate whether the effect of the observed whistler waves on electrons could be addressed within the quasi-linear theory (QLT) \citep[e.g.,][]{Sag&Gal69}. The QLT is applicable for a sufficiently wide frequency width of a whistler wave spectrum \citep[e.g.,][]{Karpman74}: $\Delta f/f_w\gg \left(B_w/B_0\right)^{1/2}\;\left(kv_{\perp}/\omega_{ce}\right)^{1/2}$, where $v_{\perp}$ is the electron velocity perpendicular to the magnetic field. Because the whistler waves interact efficiently with halo electrons, we can assume that $v_{\perp}$ is a few times larger than the electron thermal velocity. Using the cold dispersion relation for whistler waves we rewrite the QLT applicability criterion
\begin{equation}
\frac{\Delta f}{f_w}\gg \left(\frac{B_w}{B_0}\right)^{1/2}\;\left(\beta_{e}\;\frac{f_w/f_{ce}}{1-f_w/f_{ce}}\right)^{1/4}
\label{eq:qlt_appl}
\end{equation}

Figure \ref{fig:QLT_applicability} presents the test of the QLT applicability and shows that $\Delta f/f_w$ is always above the right-hand side of Eq. (\ref{eq:qlt_appl}). The inset panel shows the probability distribution function of the ratio of $\Delta f/f_w$ to the right-hand side and confirms that in the majority of the events $\Delta f/f_w$ is about five times larger than the right-hand side. We conclude that the quasi-linear theory is likely a good approximation for analysis of effects of the observed whistler waves on electrons. At the same time, we stress that an extensive statistical analysis of waveform measurements should be carried out in the future to verify that whistler wave amplitudes $B_w$ inferred from 8s magnetic field spectra do not significantly underestimate the actual \added{peak} amplitudes of whistler waves. The statement of the QLT applicability concerns only whistler waves in the pristine solar wind. Whistler waves observed in interplanetary shock waves may be rather narrow-band and large-amplitude for the QLT to be applicable \citep[e.g.,][]{Breneman10,Wilson13}. 

\begin{figure}
    \centering
    \includegraphics[width=\linewidth]{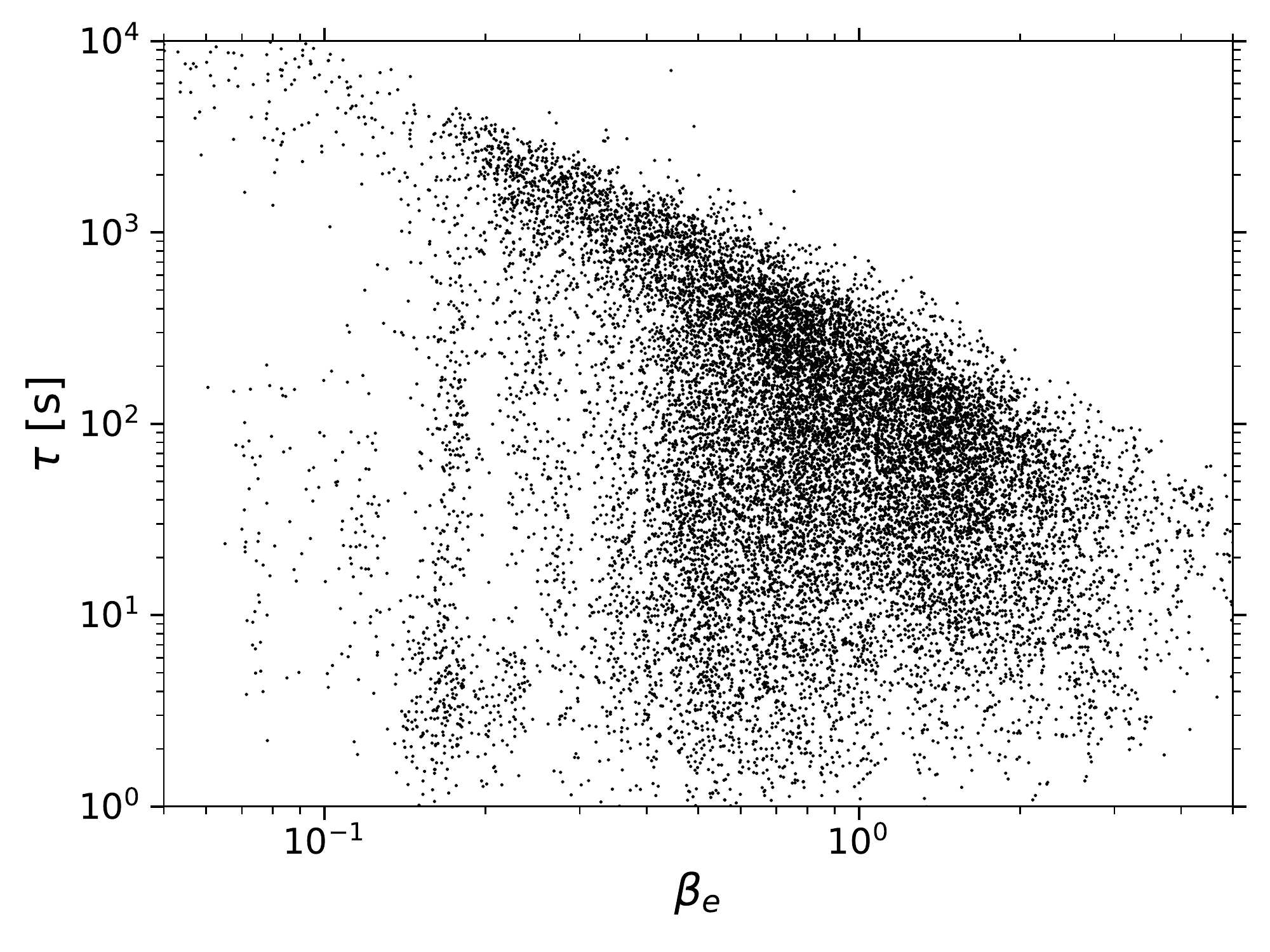}
    \caption{The quasi-linear relaxation time of unstable electron VDFs by the observed whistler waves presented versus $\beta_{e}$.}
    \label{fig:diffusion_time_vs_beta}
\end{figure}

Figure \ref{fig:diffusion_time_vs_beta} presents order of magnitude estimates of the quasi-linear relaxation time of unstable electron VDFs by the observed whistler waves. The relaxation time is given by the following expression \citep[e.g.,][]{Karpman74} 
\begin{equation}
    \tau \approx \frac{1}{2\pi f_w \beta_e} \left(\frac{\Delta f}{f_w}\right)^3 \frac{B_0^2}{B_w^2},
    \label{eq:diffusion_time}
\end{equation}
where in deriving this formula we have assumed that $f_{w}\ll f_{ce}$. The typical relaxation time is a few tens of minutes at low $\beta_{e}$ to about a minute at $\beta_{e}\sim 1$. In principle the relaxation can be as fast as a few seconds. The strong negative correlation between $\tau$ and $\beta_e$ is due to explicit dependence of $\tau$ on $\beta_{e}$ according to Eq. (\ref{eq:diffusion_time}) and due to a strong positive correlation between $B_w/B_0$ and $\beta_e$. During the relaxation time whistler waves may cover spatial distances of a few tens of thousands kilometers implying that low-frequency density and magnetic field fluctuations may affect the relaxation process of the WHFI \citep[see, e.g.,][for relaxation of a beam instability in nonuniform solar wind plasma]{Voshchepynets15}.

We have shown that the frequency \added{upper bound} of the observed whistler waves is negatively correlated with $\beta_{e}$ and demonstrated that the frequencies are in effect consistent with the theoretical predictions of the WHFI. Thus, in accordance with conclusions of \cite{Tong2019a} whistler waves observed in the pristine solar wind can be indeed generated by the WHFI operating in a local plasma. We have compared the observed frequencies to predictions of the WHFI theory with electron VDFs consisting of core and halo electron populations. The presence of the anti-sunward strahl population typical for the fast solar wind \citep{Pilipp87,Stverak09} would not affect any characteristics of the WHFI, because whistler waves produced by the WHFI propagate anti-sunward and do not resonate with the strahl \citep[e.g.,][for discussion]{Vasko2019a}. 

The original WHFI theory assumed both core and halo electron populations to be temperature isotropic \citep{Gary75}. The unstable whistler waves were shown to propagate parallel to the halo bulk velocity or, equivalently, parallel to the electron heat flux. In the realistic solar wind both core and halo populations exhibit some temperature anisotropies \citep[][]{Feldman76,Stverak08,Pierrard16}. Even a slight temperature anisotropy $T_{h\perp}/T_{h||}>1$ of the halo population increases the growth rate of whistler waves propagating parallel to the heat flux \citep[][]{Tong2019a}. At sufficiently high $T_{h\perp}/T_{h||}>1$ whistler waves propagating anti-parallel to the electron heat flux can be unstable as expected for the classical temperature anisotropy instability (TAI) \citep{Sagdeev60,Kennel66,Gary12}, which may drive whistler waves propagating both parallel and anti-parallel to the electron heat flux at any negligible or zero heat flux value. We cannot rule out that some of the whistler waves in the solar wind are driven by the TAI of the halo population and propagate opposite to the electron heat flux. At the moment, we can point out that our statistical results support the WHFI scenario, so that the major part of the whistler waves in our dataset is likely produced by the WHFI. The analysis of the TAI in the solar wind requires very careful fitting of electron VDFs and analysis of whistler waveforms (not available continuously) that is left for future studies.

Finally, we notice that whistler waves considered in this paper are electromagnetic waves that have been identified in the magnetic field spectra. We have definitely missed electrostatic whistler waves potentially present in the solar wind \citep{Vasko2019a}, but not visible in the magnetic field spectra. The results of this statistical study will be useful for the future analysis of whistler wave origin and effects, e.g., heat flux regulation and supratheramal electron scattering, in the solar wind.

\section{Conclusion \label{sec7}}

In this section we summarize the results of our statistiscal analysis of whistler waves at 1 AU:

\begin{enumerate}

    \item The intense wave activity in the whistler frequency range is shown to be dominated (80\%) by quasi-parallel whistler waves. The overall occurrence of quasi-parallel whistler waves in the pristine solar wind is found to be about 1.7\%. We emphasize that only intense whistler waves above 16 Hz have been considered in this study, so that the actual occurrence is certainly higher. 

    \item The occurrence probability of whistler waves in the pristine solar wind is strongly dependent on the electron temperature anisotropy $T_{e\perp}/T_{e||}$. The occurrence probability is less than 2\% at $T_{e\perp}/T_{e||}\lesssim 0.9$, but varies from 5 to 15\% as $T_{e\perp}/T_{e||}$ increases from 0.95 to 1.2. There is no apparent dependence of the whistler wave occurrence on the electron heat flux $q_e/q_0$, the solar wind velocity $v_{sw}$ or $\beta_{e}$.
    
    \item Whistler waves in the solar wind have amplitudes typically below 0.02 $B_0$, where $B_0$ is the magnitude of the quasi-static magnetic field. In physical units the amplitudes are in the range from about 0.01 to 0.1 nT. 
    
    \item The average normalized whistler wave amplitude $B_w/B_0$ correlates positively with $q_e/q_0$ and $T_{e\perp}/T_{e||}$, but the strongest positive correlation is found with $\beta_{e}$. The variation of $q_{e}/q_0$ and $T_{e\perp}/T_{e||}$ over the observed range results in variation of $B_w/B_0$ by \replaced{a factor of two}{about 30\%}, while variation of $\beta_{e}$ from 0.1 to 5 results in variation of $B_w/B_0$ by a factor of \replaced{four}{three}. The whistler wave amplitude negatively correlates with the solar wind velocity, varying by a factor of two from slow to fast solar wind.
    
    \item Whistler wave frequencies $f_w/f_{ce}$ \deleted{are shown to be negatively correlated with $\beta_{e}$ and} fall between some upper and lower bounds dependent on $\beta_{e}$. The upper bound on the whistler wave frequency is approximately given by $0.24\;\beta_{e}^{-0.31}$. The frequency bandwidth $\Delta f$ of the whistler waves is determined and $\Delta f/f_{w}$ is shown to be positively correlated with $\beta_{e}$.
        
    \item We show that the observed whistler wave frequencies are consistent with the theoretical predictions of the whistler heat flux instability, indicating thereby that whistler waves in the pristine solar wind can be generated by the WHFI. The generation of some of the whistler waves by the temperature anisotropy instability can not be ruled out.
    
    \item We have shown that the frequency width of the whistler waves is sufficiently wide so that the quasi-linear theory is likely applicable to describe effects of the whistler waves on electrons. The typical quasi-linear relaxation time in a uniform plasma would be from a minute at $\beta_{e}\sim 1$ to a few tens of minutes at low $\beta_{e}$. In principle the relaxation can be as fast as a few seconds.
    
    \item We have estimated the energies of electrons resonating the whistler waves and shown that the whistler waves should be driven by suprathemral electrons, whose minimum energy $E_R$ is negatively correlated with $\beta_{e}$. \replaced{$E_R$ is a few tens of eV at $\beta_{e}\sim 1$ and above a few hundred eV at $\beta_e\sim 0.1$.}{$E_R$ is about a few tens of eV, or equivalently, about three times the thermal energy at $\beta_e\sim 1$, and about a few hundred eV or about ten times the thermal energy at at low $\beta_{e}$.}
\end{enumerate}

\acknowledgments
We acknowledge the THEMIS team for the use of data. The initial data access and processing was done using SPEDAS V3.1 \citep{Angelopoulos19}. We thank Trevor A. Bowen, Marc Pulupa, Vladimir Krasnoselskikh and Lynn B. Wilson III for useful discussions. Y. T. and S. D. B. were supported in part by NASA contract NNN06AA01C.

\end{document}